\documentclass[twocolumn,superscriptaddress,prd]{revtex4-1} 

\pdfoutput=1 

\usepackage{verbatim}
\usepackage{mathtools}
\usepackage{relsize}
\usepackage{natbib}

\usepackage{amsmath}
\usepackage{graphicx}
\usepackage[capposition=bottom]{floatrow}
\usepackage{amssymb}
\usepackage{graphicx}
\usepackage{rotating}
\usepackage{afterpage}
\usepackage{float}
\usepackage{geometry}
\usepackage{titlesec}
\usepackage{epstopdf}

\usepackage{latexsym}
\usepackage{graphicx}
\usepackage{color}
\usepackage{array}
\usepackage{multirow}
\usepackage{wasysym}
\usepackage{verbatim}
\usepackage{subfig}

\newcommand{\isot}[2]{$^{#2}$#1 }

\newcommand{\krsrc}{\isot{Kr}{83m}}
\newcommand{\cssrc}{\isot{Cs}{137}}

\renewcommand\thesection{\Roman{section}.}
\renewcommand\thesubsection{\arabic{subsection}.}
\renewcommand\thesubsubsection{\alph{subsubsection}.}
\titleformat{\section}[block]{\bfseries\filcenter}{\thesection}{1em}{}
\titleformat{\subsection}[block]{\bfseries\filcenter}{\thesubsection}{1em}{}
\titleformat{\subsubsection}{\bfseries\filcenter}{\thesubsubsection}{1em}{}

\usepackage{titlesec}

\begin{document}

\title{Tritium calibration of the LUX dark matter experiment}

\author{D.S.~Akerib} 
\affiliation{Case Western Reserve University, Department of Physics, 10900 Euclid Ave, Cleveland, OH 44106, USA}
\affiliation{SLAC National Accelerator Laboratory, 2575 Sand Hill Road, Menlo Park, CA 94205, USA}
\affiliation{Kavli Institute for Particle Astrophysics and Cosmology, Stanford University, 452 Lomita Mall, Stanford, CA 94309, USA}

\author{H.M.~Ara\'{u}jo} 
\affiliation{Imperial College London, High Energy Physics, Blackett Laboratory, London SW7 2BZ, United Kingdom}

\author{X.~Bai} 
\affiliation{South Dakota School of Mines and Technology, 501 East St Joseph St., Rapid City, SD 57701, USA}

\author{A.J.~Bailey} 
\affiliation{Imperial College London, High Energy Physics, Blackett Laboratory, London SW7 2BZ, United Kingdom}

\author{J.~Balajthy} 
\affiliation{University of Maryland, Department of Physics, College Park, MD 20742, USA}


\author{P.~Beltrame} 
\affiliation{SUPA, School of Physics and Astronomy, University of Edinburgh, Edinburgh EH9 3FD, United Kingdom}

\author{E.P.~Bernard} 
\affiliation{Yale University, Department of Physics, 217 Prospect St., New Haven, CT 06511, USA}

\author{A.~Bernstein} 
\affiliation{Lawrence Livermore National Laboratory, 7000 East Ave., Livermore, CA 94551, USA}

\author{T.P.~Biesiadzinski} 
\affiliation{Case Western Reserve University, Department of Physics, 10900 Euclid Ave, Cleveland, OH 44106, USA}
\affiliation{SLAC National Accelerator Laboratory, 2575 Sand Hill Road, Menlo Park, CA 94205, USA}
\affiliation{Kavli Institute for Particle Astrophysics and Cosmology, Stanford University, 452 Lomita Mall, Stanford, CA 94309, USA}


\author{E.M.~Boulton} 
\affiliation{Yale University, Department of Physics, 217 Prospect St., New Haven, CT 06511, USA}

% ...

 \begin{abstract}

We present measurements of the electron-recoil (ER) response of the LUX dark matter detector based upon 170,000 highly pure and spatially-uniform tritium decays. We reconstruct the tritium energy spectrum using the combined energy model and find good agreement with expectations. We report the average charge and light yields of ER events in liquid xenon at 180 V/cm and 105 V/cm and compare the results to the NEST model. We also measure the mean charge recombination fraction and its fluctuations, and we investigate the location and width of the LUX ER band. These results provide input to a re-analysis of the LUX Run3 WIMP search .

\end{abstract}

		\maketitle

\section{Introduction}

The Large Underground Xenon experiment (LUX) is a WIMP search located at the 4850' level of the Sanford Underground Research Facility (SURF) in Lead, South Dakota~\cite{lux-nim}. LUX detects particle interactions in liquid xenon (LXe) via scintillation (S1) and ionization charge (S2) signals. The LXe is instrumented as a dual-phase time projection chamber (TPC), providing an energy measurement, position information in three dimensions, and single-scatter event identification. Electron-recoil (ER) and nuclear-recoil (NR) interactions are distinguished by the ratio of the charge and light signals (S2/S1). Results from the first LUX science run (Run 3) were first reported in Ref.~\cite{lux-prl}. An improved analysis of the Run 3 data is reported in Ref.~\cite{lux-reanalysis}.

To calibrate the ER response of LUX, external gamma sources such as \cssrc are occasionally employed, but such sources are unable to produce a useful rate of fiducial single-scatter events in the WIMP energy range of interest due to self-shielding. Therefore the ER response is monitored and calibrated primarily with electron-emitting radioisotopes that can be dissolved in the LXe. Two such sources,  \krsrc~\cite{Kastens:2009pa, Baudis} and tritium ($^{3}$H), have been deployed, both providing a large sample of spatially-uniform events. In this article we report results from the calibration of LUX with tritium, a single-beta emitter with a Q-value of 18.6~keV electron-equivalent\footnote{ER events and NR events generally have different energy scales in LXe. In this article we interpret all events using the ER energy scale.}~\cite{Tritium_Q}. Neutron sources and a neutron generator are also employed by LUX to study the response to NR events~\cite{lux-reanalysis}.

The tritium beta spectrum is well known both theoretically and experimentally. It has a broad peak at 2.5~keV and a mean energy of 5.6~keV~\cite{Tritium_Mean,Tritium_Eq,Drexlin:2013lha}. 64.2\% of the decays occur between 1 and 8~keV, the energy range of interest for WIMP searches in LUX. These characteristics make it an ideal source for studying the ER response of the detector.  $\rm ^{83m}$Kr, which emits 9.4~keV and 32.1~keV internal conversion electrons, is well suited for routine monitoring and for correcting the spatial and temporal variations of the S1 and S2 signals, but is less useful for studies of the S2/S1 ER discrimination variable because both conversion electrons are above the dark matter energy range, and because the S2 signals from the two electrons generally overlap in the detector due to the short half-life of the intermediate state (154~ns). We note that the most important background in LUX is due to Compton scatters, and such events are expected to have similar properties to beta decays in the tritium energy range~\cite{NEST_2013}. 

We use tritiated methane (CH$_3$T) as the host molecule to deliver tritium activity into LUX. Compared to molecular tritium (T$_2$), CH$_3$T has several advantages. It does not adsorb onto surfaces like the T$_2$ molecule, and it does not interfere with charge transport in LXe. Also, because of its 12.3 year half-life, tritium must be removed from the detector by purification, and methane is amenable to chemical removal with standard noble gas purifiers~\cite{Dobi_CH4}. Note, however, that diffusion of tritium activity into plastic detector components during the calibration is an important concern, since that activity may later re-contaminate the LXe during the WIMP search runs.  In this respect, CH$_3$T is preferable over T$_2$ due to its larger molecular size and lower diffusion constant and solubility~\cite{miyake:1983}. We investigated the CH$_3$T contamination risk empirically with a series of bench-top tests prior to the first injection into LUX. These tests, which are described in Appendix \ref{sec:appendix1}, demonstrated that the injection and removal could be done without undue risk to the experiment. 

An initial tritium dataset of $\sim$7,000 fiducial events was obtained in August of 2013, and the results were reported in Ref.~\cite{lux-prl}. Subsequently, in December 2013, we injected additional activity with a higher rate and obtained a fiducial tritium dataset of 170,000 events. This dataset is used to characterize the LUX ER band in Ref.~\cite{lux-reanalysis}. Except where otherwise noted, in this article we report results from the larger December 2013 dataset.

\section{Injection and removal of CH$_3$T}

Two CH$_3$T sources with total activities of 3 Bq and 200 Bq were prepared for use in LUX. Each source is contained in a 2.25 liter stainless steel bottle and is mixed with 2 atmospheres of LUX-quality purified xenon. The xenon acts as a carrier gas to extract the source from the bottle. The CH$_3$T was synthesized by Moravek Biochemical~\cite{moravek} and delivered at a specific activity of 0.1 milliCurie per millimol.

\begin{figure}[t!]
\includegraphics[width=80mm]{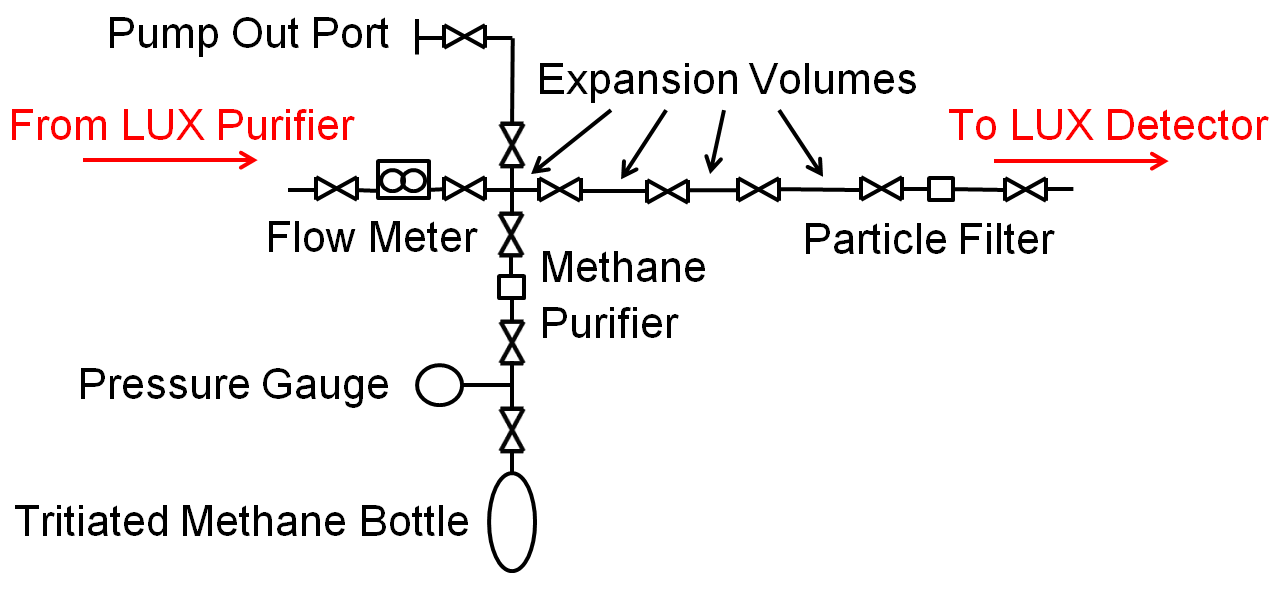}
\caption{Plumbing diagram of the CH$_3$T injection system for LUX. CH$_3$T is injected downstream of the xenon gas purifier so that it passes through the detector prior to being removed.  Red arrows indicate the direction of flow.}
\label{fig:plumbing}
\end{figure}

The injection system is shown in Fig.~\ref{fig:plumbing}. A fraction of the source bottle activity may be extracted by allowing the carrier gas to expand into one or more expansion volumes consisting of various sections of evacuated tubing. The amount of extracted activity is controlled by selecting an expansion volume of appropriate size. A methane purifier (SAES model MC1-905F\cite{seas}) located between the source bottle and the expansion volume ensures that only CH$_3$T, CH$_4$, and noble gases are allowed to enter the system. The extracted activity is then injected into the TPC by diverting a small portion of the LUX xenon gas flow through the expansion volumes. 

The CH$_3$T appears in the TPC within minutes of the injection, and is removed via the normal action of the LUX xenon purification system, which operates without interruption during the entire procedure. Its centerpiece is a hot zirconium getter (SAES model PS4-MT15-R1\cite{seas}) that acts upon gaseous xenon and continuously removes all non-noble species including methane. The xenon gas flow is driven by a diaphragm pump at a rate of $\sim$27 standard liters per minute (slpm).

\begin{figure}[t!]
\includegraphics[width=90mm]{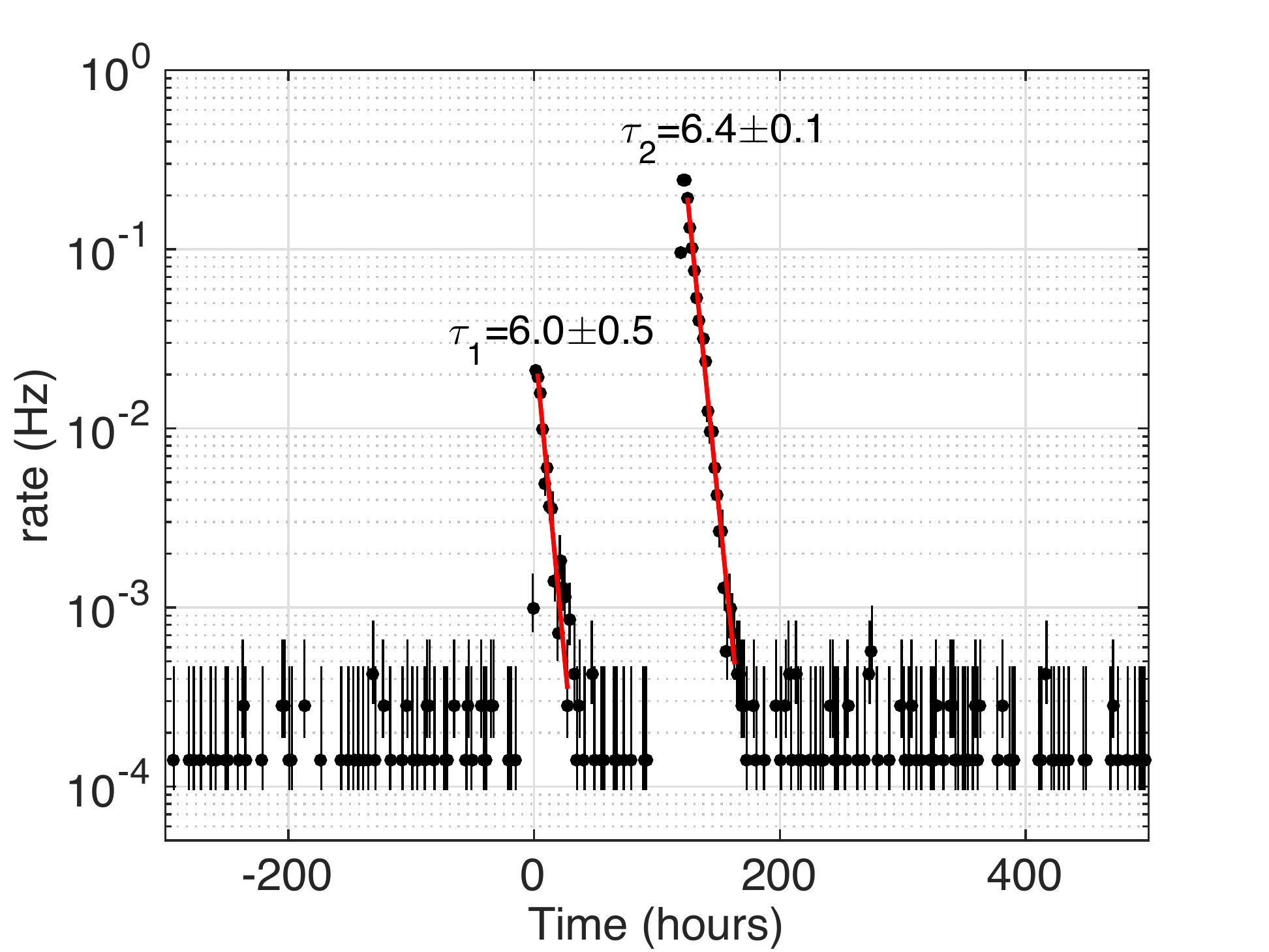}
\caption{Rate of single scatter events with S1 below 150 phd in the fiducial volume during the August 2013 CH$_3$T injections.  The solid lines are exponential fits to the activity vs. time.}
\label{fig:ch3t_removal}
\end{figure}

\begin{figure}[t!]
\includegraphics[width=90mm]{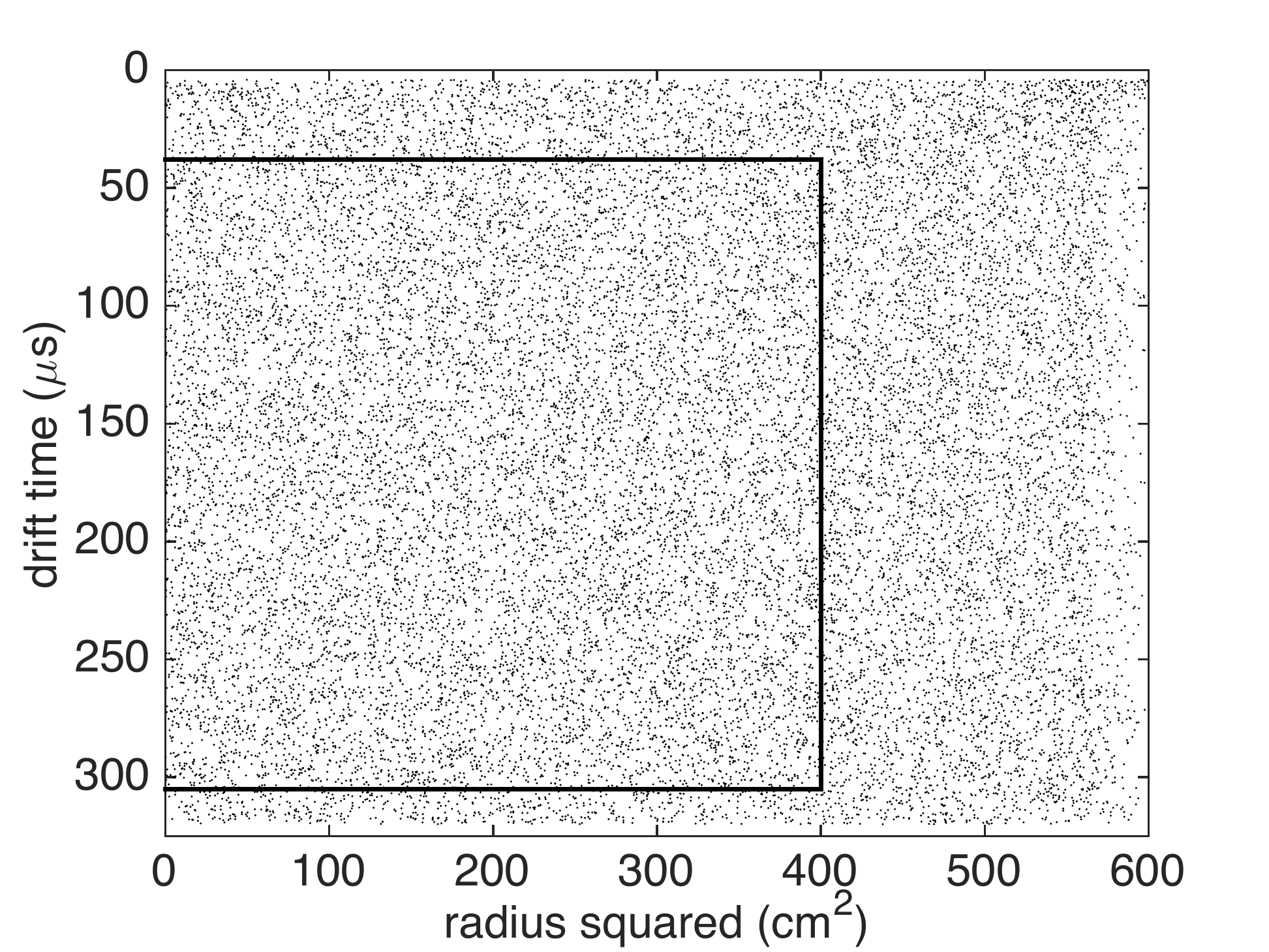}
\caption{The location of events in drift time vs. detector radius squared for the August 2013 CH$_3$T injection. The drift time is a proxy for the $z$ coordinate of the event. The solid black line represents the fiducial volume used in Ref.~\cite{lux-reanalysis}.}
\label{fig:event_location}
\end{figure}

Prior to the first injection of CH$_3$T activity, we first confirmed that the LUX getter unit was capable of efficient methane removal by injecting  $\sim$1 ppm (part-per-million g/g) of natural methane (CH$_4$) into LUX. As shown in Appendix~\ref{sec:appendix1}, the CH$_4$ concentration in the gas, monitored with a mass spectrometer, was observed  to decrease exponentially with a time constant of 5.9 $\pm 0.07$ hours. The one-pass efficiency of the getter for CH$_4$ removal was measured to be 97\% under the LUX flow and temperature conditions by sampling the gas before and after the getter. 

On August 8, 2013, an initial injection of 20 mBq of CH$_3$T was performed, followed five days later by an injection of 800 mBq. The count rate of fiducial single-scatter events with S1 $<$ 150 photons detected (phd) (roughly the endpoint of the tritium beta spectrum) is shown in Fig.~\ref{fig:ch3t_removal}. The CH$_3$T activity is clearly observed, with the count rate reaching its maximal value in one hour. For both injections the activity was removed with a six-hour exponential time constant similar to that observed in the CH$_4$ injection.The location of the CH$_3$T events from the first injection after all corrections is shown in Fig.~\ref{fig:event_location}. As expected, the events are uniform within the detector volume.  

It is worth noting that the observed purification time constant is considerably shorter 
than the xenon mass turn-over time of LUX (about 40 hours for 370 kg of xenon).  The 
LUX purification circuit is somewhat complex, including both LXe flow drawn from the top of 
the detector as well as gas flow drawn from the anode region.  A simple and descriptive 
model of LUX purification is presented in Appendix~\ref{sec:appendix2}. A more 
complete study of LUX purification that addresses the physical origin of the short purification
time is not possible with the present data.

\section{Results}

At the conclusion of Run 3, in December of 2013, a total of 10 Bq of tritium was injected into LUX and removed. 300,000 events were observed in the 250 kg active volume, of which 170,000 events were in the 145~kg fiducial volume at the nominal LUX electric field of 180~V/cm. Another 4,500 fiducial events were collected in a special run at a reduced field of 105~V/cm. 

The LUX detector is described in detail in Ref.~\cite{lux-nim}. Briefly, LUX is a cylindrical dual-phase TPC, with an array of 61 photomultiplier tubes (PMTs) immersed in the LXe at the bottom of the vessel, and an identical PMT array above the liquid-gas interface. Primary scintillation signals (S1) are detected on both arrays, while ionization electrons drift vertically in the uniform drift field as established by anode and cathode wire grids. The ionization charge is extracted through the liquid-gas surface and creates secondary scintillation (S2) before being collected by the anode. The S2 signal is detected by both arrays, and its spatial pattern on the upper array localizes the event in $x$ and $y$. The time between S1 and S2 determines the $z$ coordinate.

Data are selected for analysis using cuts similar to those employed in the WIMP search analysis\cite{lux-reanalysis, lux-prd}. Within an event window, single scatters are selected by pairing an S1 with a single S2.  The S1 is measured with a spike-counting method that requires a minimum two-fold coincidence from PMTs that are not in neighboring channels. We correct the S1 and S2 signals for spatial and temporal variations such as the light collection efficiency and the free electron lifetime with $ ^{83m}$Kr data.  We report the S1 and 
S2 signal sizes in units of photons detected (phd)~\cite{lux-reanalysis}, a measure which more accurately reflects the true number of VUV quanta compared to the more familiar photoelectron counting by properly accounting for double photoelectron emission as reported in Ref.~\cite{doublepe}. The S2 signal is required to be greater than 165 phd ($\sim$6 extracted electrons) to ensure accurate $x$-$y$ position reconstruction. Events are required to be within a fiducial volume between 38 and 305~$\rm \mu s$ in drift time (8.5 and 48.6 cm in the charge drift direction ($z$) measured from the face of the bottom PMTs) and less than 20~cm radius. In addition to the above selection cuts, which are applied to the WIMP search, in the tritium data we also reject events where the S2 signal is truncated by the end of an event buffer. This pathology is negligible in WIMP search data but is present at a small level in the tritium data due to the larger event rate.

\subsection{Tritium energy spectrum}

We interpret the data in terms of the combined energy model for electron recoils~\cite{Platzman}, where the total energy of an interaction is directly proportional to the number of quanta produced (ionization electrons plus scintillation photons):

\begin{equation}
E_{total} = W \cdot (n_{\gamma} + n_e ),
\label{platzman_eq}
\end{equation}

\noindent
where $E_{total}$ is the energy of the deposition in keV and  $n_\gamma$ and $n_e$ are the number of photons and electrons, respectively. We employ the combined energy model because it reproduces well the true energy of the event, while the individual photon and electron signals are non-linear in energy due to the effects of recombination. We use a $W$-value of 13.7 $\pm$ 0.2 eV/quantum~\cite{Dahl_Thesis}. In LUX $n_{\gamma}$ and $n_e$ are proportional to the S1 and S2 signals, with gain factors $g_1$ and $g_2$: 

\begin{equation}
E_{total} = W \cdot \left(\frac{S1}{g_1} + \frac{S2}{g_2} \right),
\label{energy_eq}
\end{equation}

\noindent
where S1 and S2 have units of phd and $g_1$ and $g_2$ have units of phd/quantum. $g_1$ is the light collection efficiency referenced to the center of the detector times the average quantum efficiency of the PMT arrays, while $g_2$ is the product of the electron extraction efficiency at the liquid-gas interface and the average size of the single electron response in phd. For the December 2013 tritium dataset presented here, $g_1$,$g_2$, and the extraction efficiency are measured to be $0.115 \pm 0.005$ phd/photon, $12.1 \pm 0.9$ phd/electron, and $50.9\% \pm 3.8\%$. The constraint was set by allowing $g_1$ and $g_2$ to float and fitting the data to a true tritium spectrum~\cite{Drexlin:2013lha}.  In the LUX Run 3 WIMP search, $g_1$, $g_2$, and the extraction efficiency are measured with mono-energetic source data and single electron events to be $0.117 \pm 0.003$ phd/photon, $12.1 \pm 0.8$ phd/electron, and $49.1\% \pm 3.2\%$~\cite{lux-reanalysis, lux-prd}, consistent with the values adopted here. The value of $g_1$ is also consistent with expectations from a Monte Carlo simulation of LUX~\cite{lux-prd}, while the value of the electron extraction efficiency is consistent with bench-top measurements~\cite{gushchin:1979,gushchin:1982}. The consistency $g_1$ and $g_2$ with expectations provides evidence that the $W$-value adopted here is valid for the tritium energy range, although an exact determination of $W$ is not possible from this data. 

\begin{figure}[t!]
\includegraphics[width=90mm]{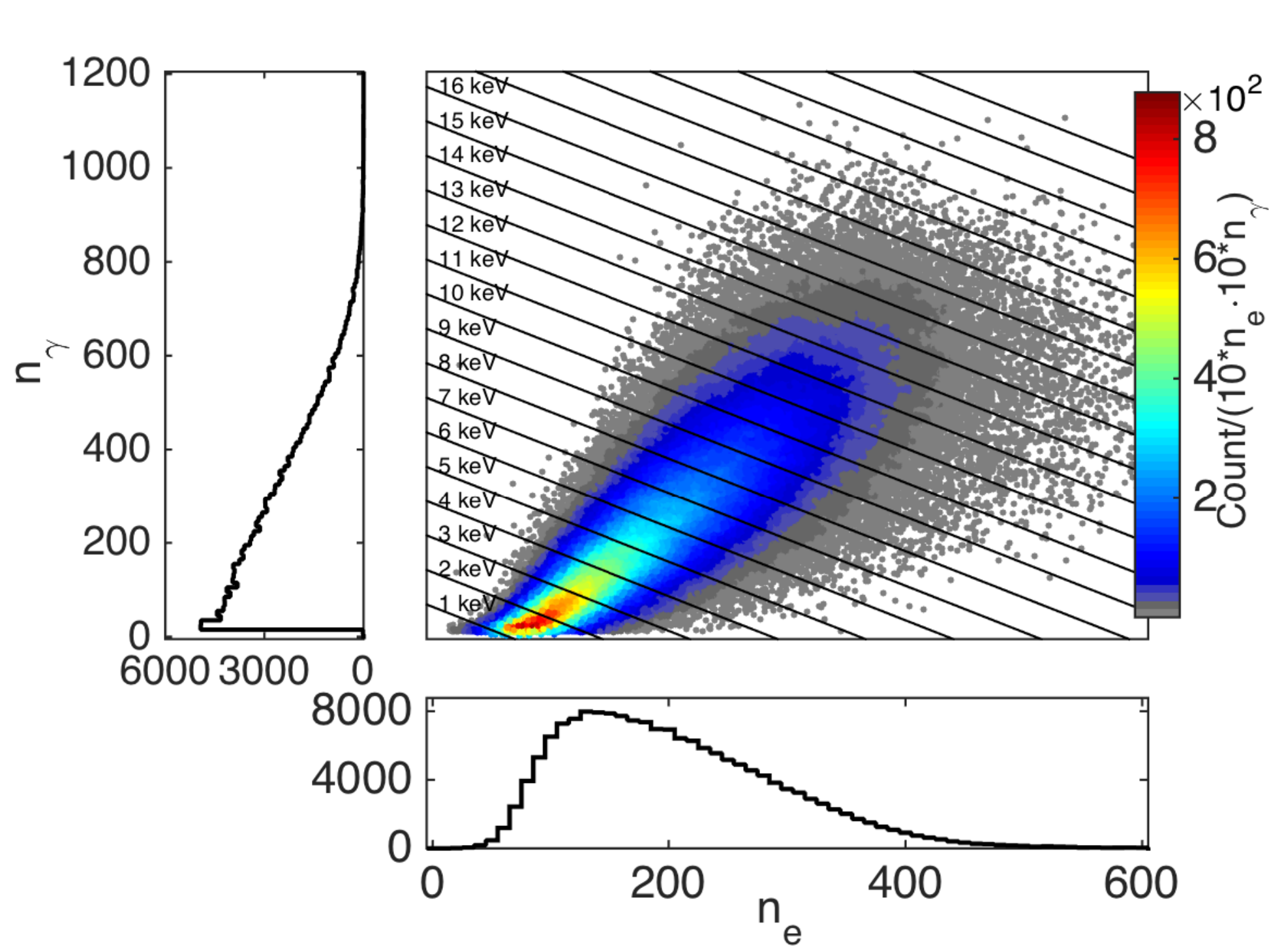}
\caption{Scatter plot of $n_e$ vs $n_{\gamma}$ for 170,000 fiducial tritium events at 180~V/cm. Lines of constant energy are indicated assuming a $W$-value of 13.7~eV. The data are projected onto $n_e$ and $n_{\gamma}$ histograms on each axis.}
\label{fig:tritium-scatter}
\end{figure}

\begin{figure}[t!]
\begin{center}
\includegraphics[width=90mm]{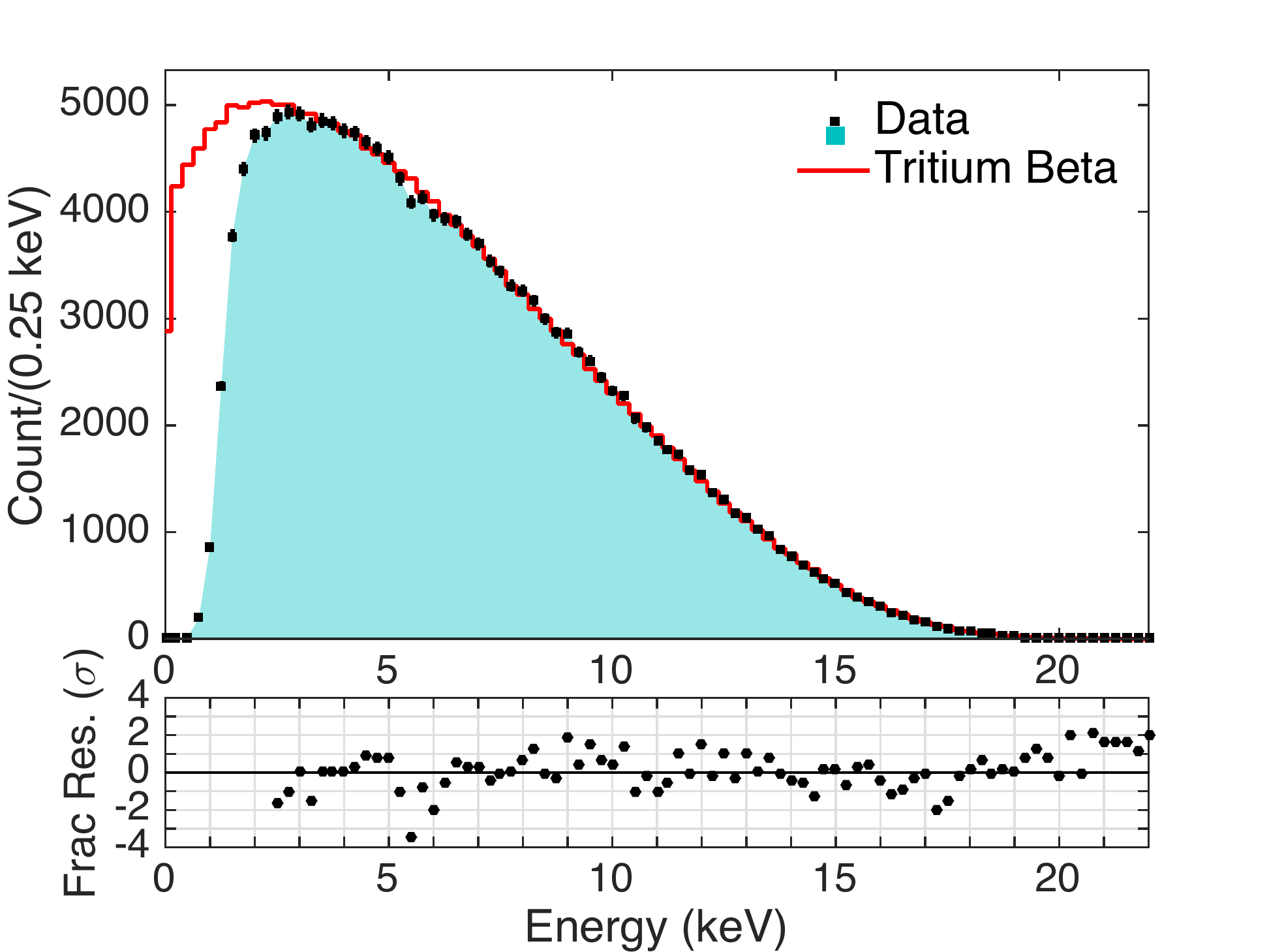}
\caption{Top: The tritium energy spectrum measured by LUX with the combined energy model (black) compared to  a tritium spectrum convolved with detector resolution  ($\frac{\sigma_E}{W} = \sqrt{\sigma^2(n_{\gamma})+ \sigma^2(n_e)}$.) The p-value between data and model from 3 to 18 keV is 0.70. Bottom: Bin-by-bin fit residuals between data and theory, in units of $\sigma$. }
\label{fig:tritium-spectrum}
\end{center}
\end{figure}

A scatter plot of $n_e$ vs $n_{\gamma}$ for the tritium data at 180~V/cm is shown in Fig.~\ref{fig:tritium-scatter}, along with the projected histograms on each axis. Contours of constant energy in 1~keV intervals are also plotted, derived from Eq.~\ref{platzman_eq}. 

\begin{figure}[t!]
\includegraphics[width=90mm]{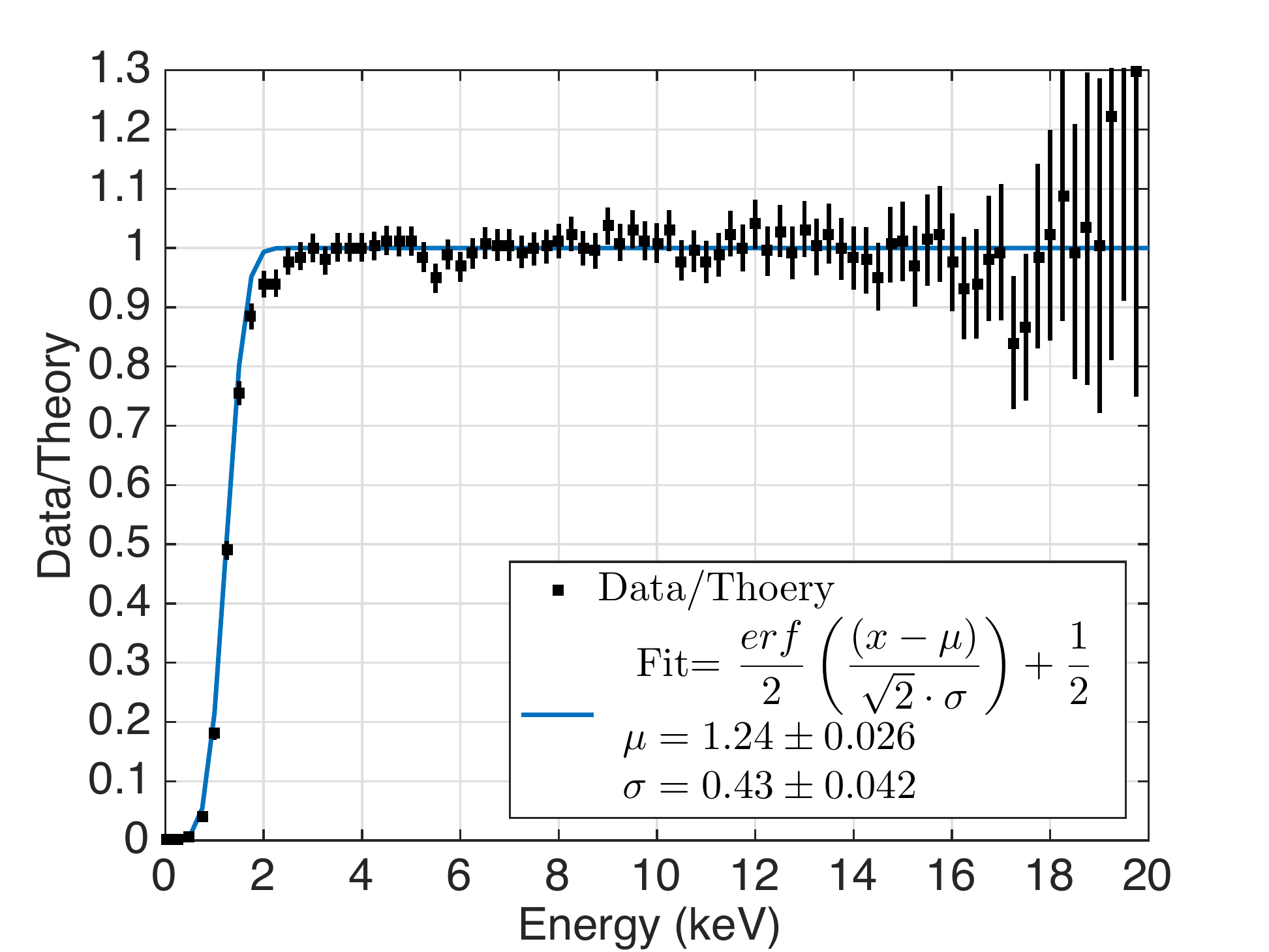}
\caption{Ratio of the measured tritium energy spectrum and the true one convolved with the detector resolution. A fit to an error function is shown.}
\label{fig:ER-threshold}
\end{figure}

The tritium energy spectrum, obtained by projecting the data along the lines of constant energy, is shown in Fig.~\ref{fig:tritium-spectrum}. The data are compared to a tritium spectrum with an applied energy resolution of $ \sigma_E = W \cdot \sqrt{\sigma(n_{\gamma})^2 + \sigma(n_e)^2}$, where $ \sigma(n_{\gamma})$ and $ \sigma(n_e)$ represent the detector resolution for photon and electron counting. In the fit the model is normalized to the data. The ratio of the data to the smeared theoretical spectrum is shown in Fig.~\ref{fig:ER-threshold}, along with an empirical fit to an error function. The effective 50\% energy threshold for ER events is found to be 1.24 $\pm$ 0.026~keV. The excellent agreement between data and theory from 3~keV to the endpoint of the tritium spectrum provides powerful support for the combined energy model of Eq.~\ref{platzman_eq}.

\subsection{Light and charge yields}

\begin{figure}[t!]
\includegraphics[width=90mm]{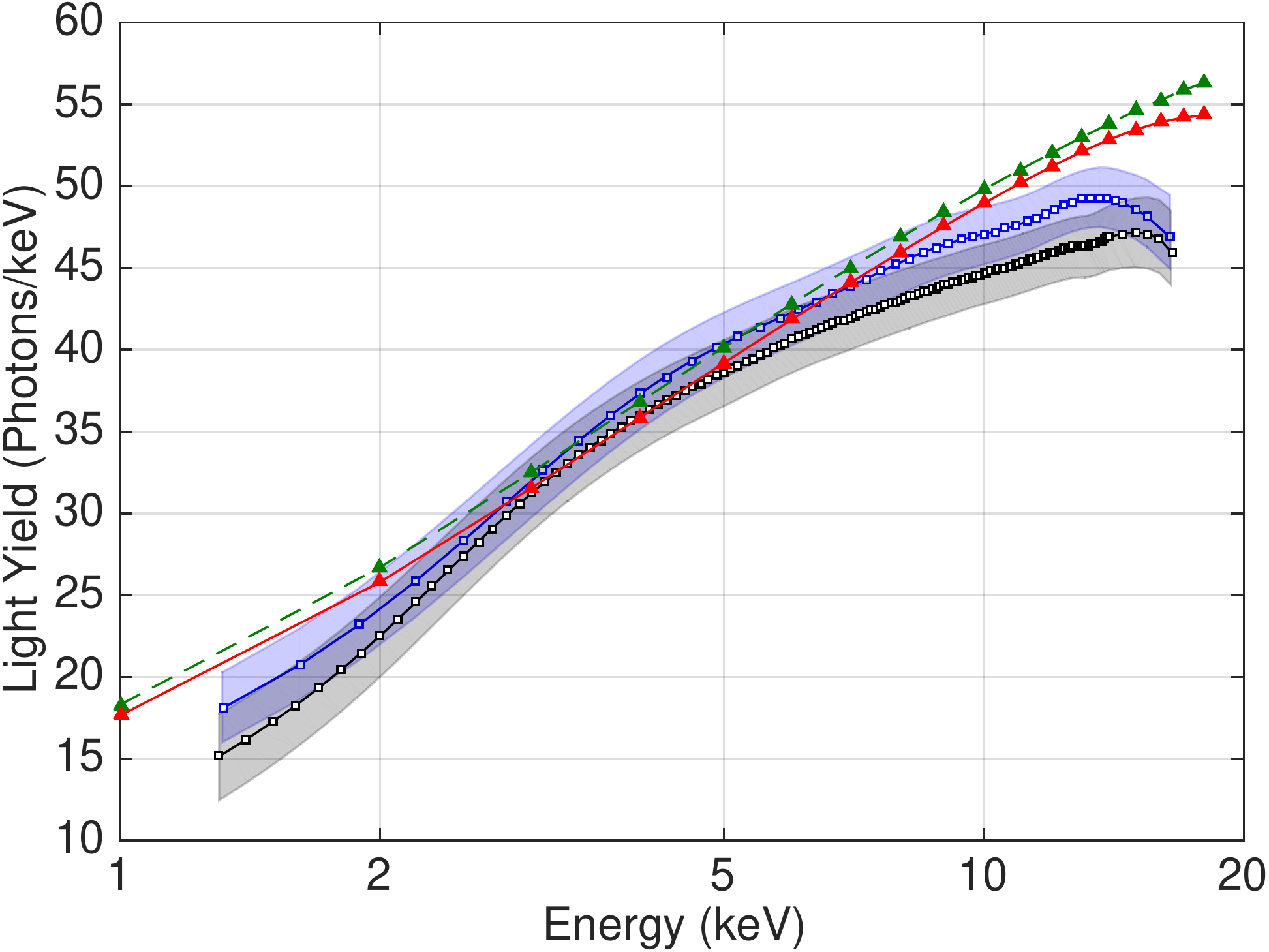}
\includegraphics[width=90mm]{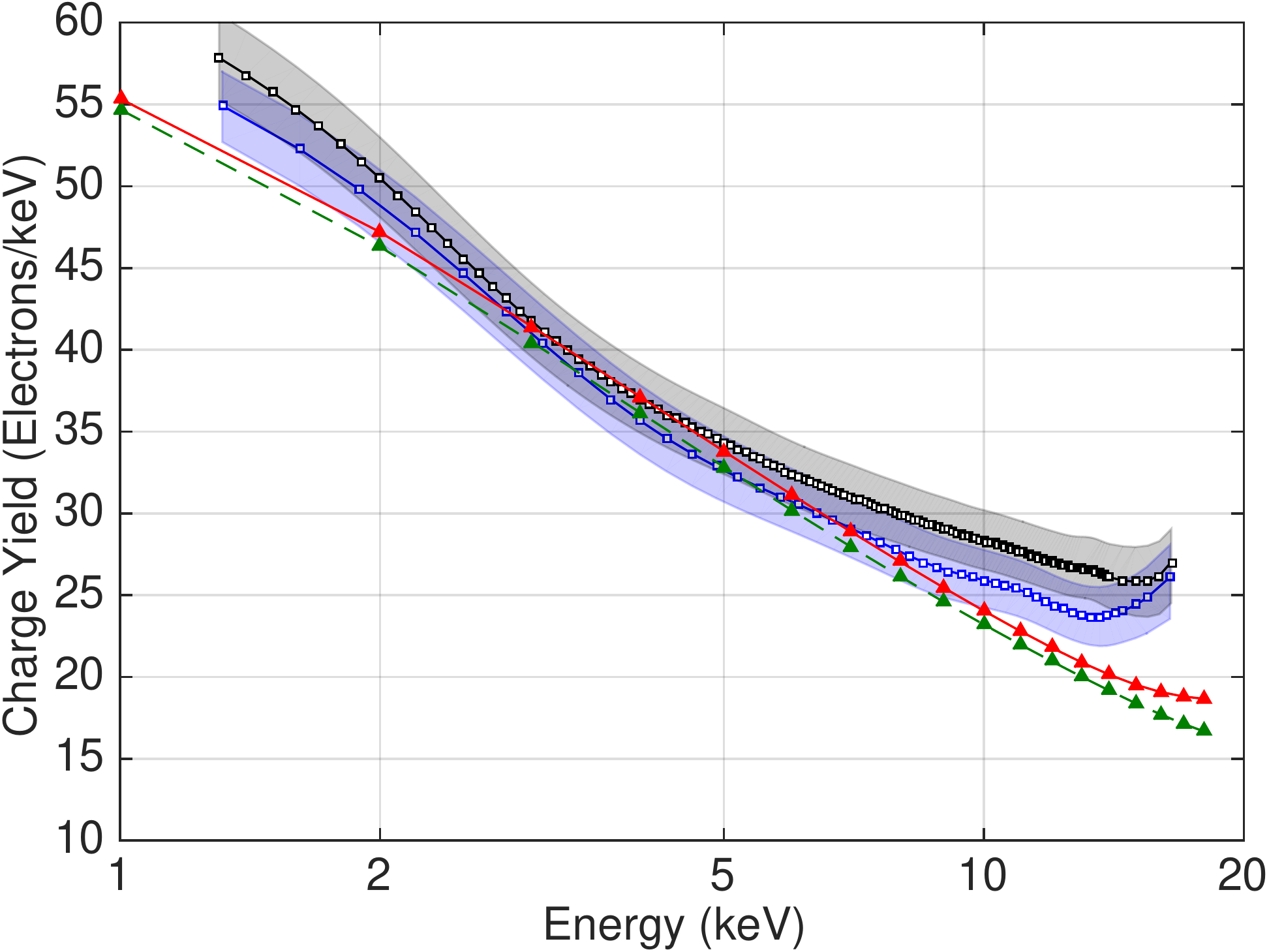}
\caption{The light yield (upper plot) and charge yield (lower plot) of tritium ER events in LUX at 180~V/cm (black squares) and 105~V/cm (blue squares) compared to NEST v0.98 (2013)~\cite{NEST_2013}. The NEST curves are solid red and dashed green for 180 and 105 V/cm respectively, with triangle markers spaced every one keV. The bands indicate the $1\sigma$ systematic uncertainties on the data due to $g_1$ and $g_2$, which are fully anti-correlated between the charge yield and light yield across all energy bins. Statistical uncertainties are negligible in comparison.}
\label{fig:ER-LY-QY}
\end{figure}

The mean light and charge yields of ER events in LUX are obtained by dividing the mean light and charge signals by the combined energy in each energy bin. The result is shown for 180~V/cm and 105~V/cm in Fig.~\ref{fig:ER-LY-QY}, along with NEST v0.98 model predictions at each field~\cite{NEST_2013}. For these plots a small correction has been applied to the data to account for smearing of tritium events across energy bins due to the energy resolution and the spectral shape~\cite{Dobi_Thesis}\footnote{We have verified with an internal $\rm ^{83m}Kr$ calibration source that the light yield of LXe is unaffected by the presence of CH$_4$ at concentrations up to $\sim$1 part per million. For the CH$_3$T measurements reported here the concentration was ($<$ $10\times10^{-12}$ g/g). }.  NEST v0.98 describes the data approximately, but predicts too much light yield and too little charge yield above 6~keV. Note that NEST v0.98 lacks direct input measurements in this energy range and electric field, so a modest disagreement is not unexpected. A version of NEST tuned to reproduce the LUX tritium data faithfully is used to model the ER response in the Run 3 re-analysis~\cite{lux-reanalysis}. The yield measurements at 180~V/cm and 105~V/cm are also listed in Tables~\ref{table:Yields} and~\ref{table:Yields_100} in Appendix~\ref{sec:appendix3}.

\begin{figure}[t!]
\includegraphics[width=88mm]{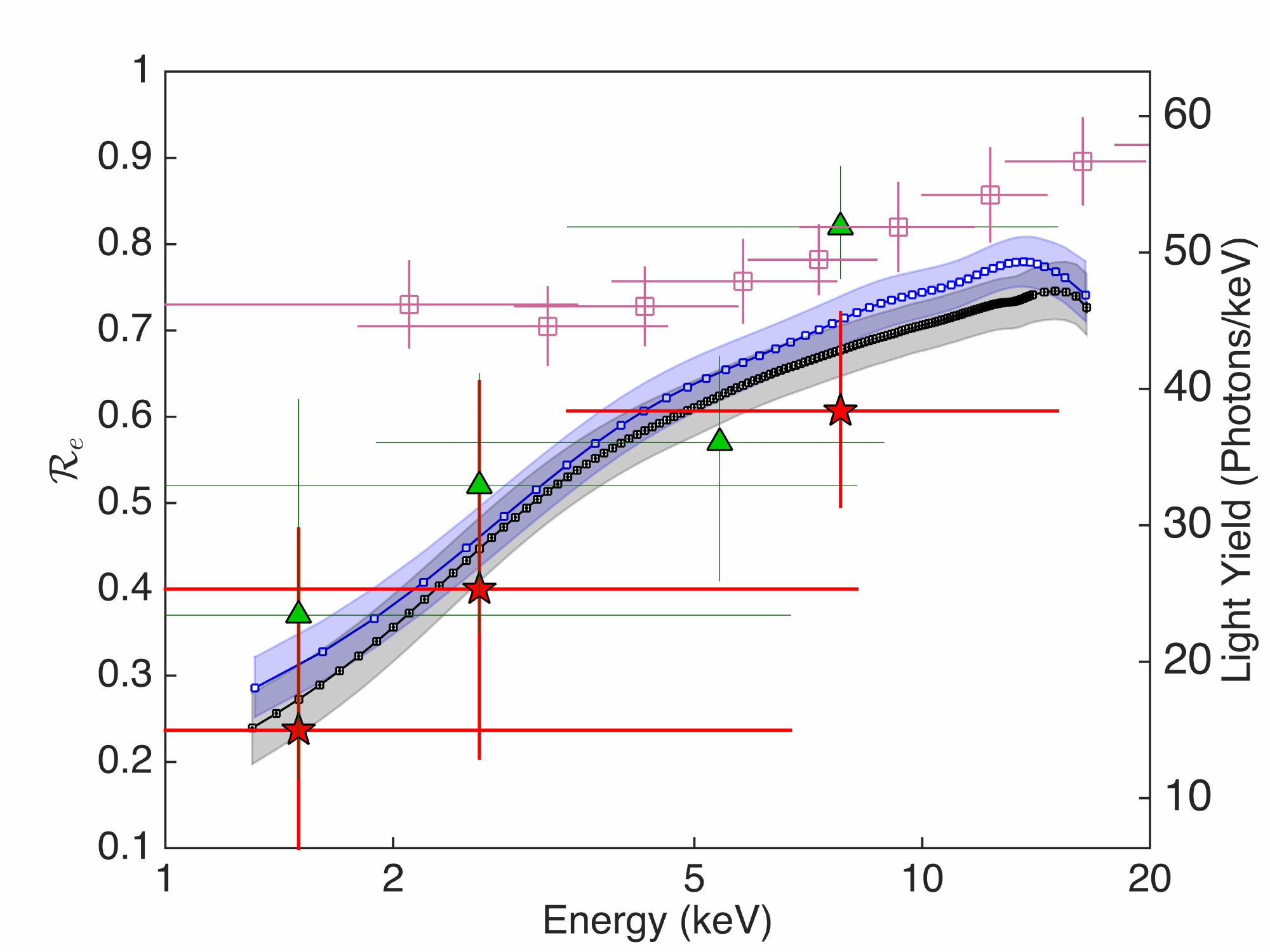}
\caption{Light yield measurement from LUX tritium data compared with results from other authors. Left vertical scale: light yield relative to that of the 32.1~keV decay of $\rm ^{83m}Kr $ at zero field. Right vertical scale: absolute light yield measurements. Blue squares represent tritium at 105~V/cm, black squares are tritium at 180~V/cm. The shaded bands are the the systematic errors on the tritium data. Magenta squares represent zero field measurements from \cite{Aprile_LY}, green triangles and red stars represent zero field and 450~V/cm from \cite{Baudis}. All non-tritium data is from Compton scatters. }
\label{fig:Re_LY}
\end{figure}

The light yield measurements are compared to similar measurements by other authors in Fig.~\ref{fig:Re_LY}. To remove detector effects from this comparison, the light yield is normalized to that of the 32.1~keV electron capture decay of $\rm ^{83m} Kr$ at zero electric field. For LUX this light yield is measured to be $ 63.3 \pm 3$ photons/keV. Although the error bars on the comparison data are large, the findings are consistent with the expectation that the light yields at 105 and 180~V/cm lie between those at zero field and 450~V/cm from Refs.~\cite{Aprile_LY}~and~\cite{Baudis}. It is worth noting that Refs.~\cite{Aprile_LY}~and~\cite{Baudis} use Compton scatters as the source of ER events, while in tritium data the ER source is a beta decay. At low energy beta particles and Compton electrons will lead to similar track lengths and are expected to produce similar event characteristics~\cite{NEST_2013}. The comparison of Fig.~\ref{fig:Re_LY} provides modest support for this expectation, albeit with large experimental uncertainties.

\begin{figure}[t!]
\includegraphics[width=90mm]{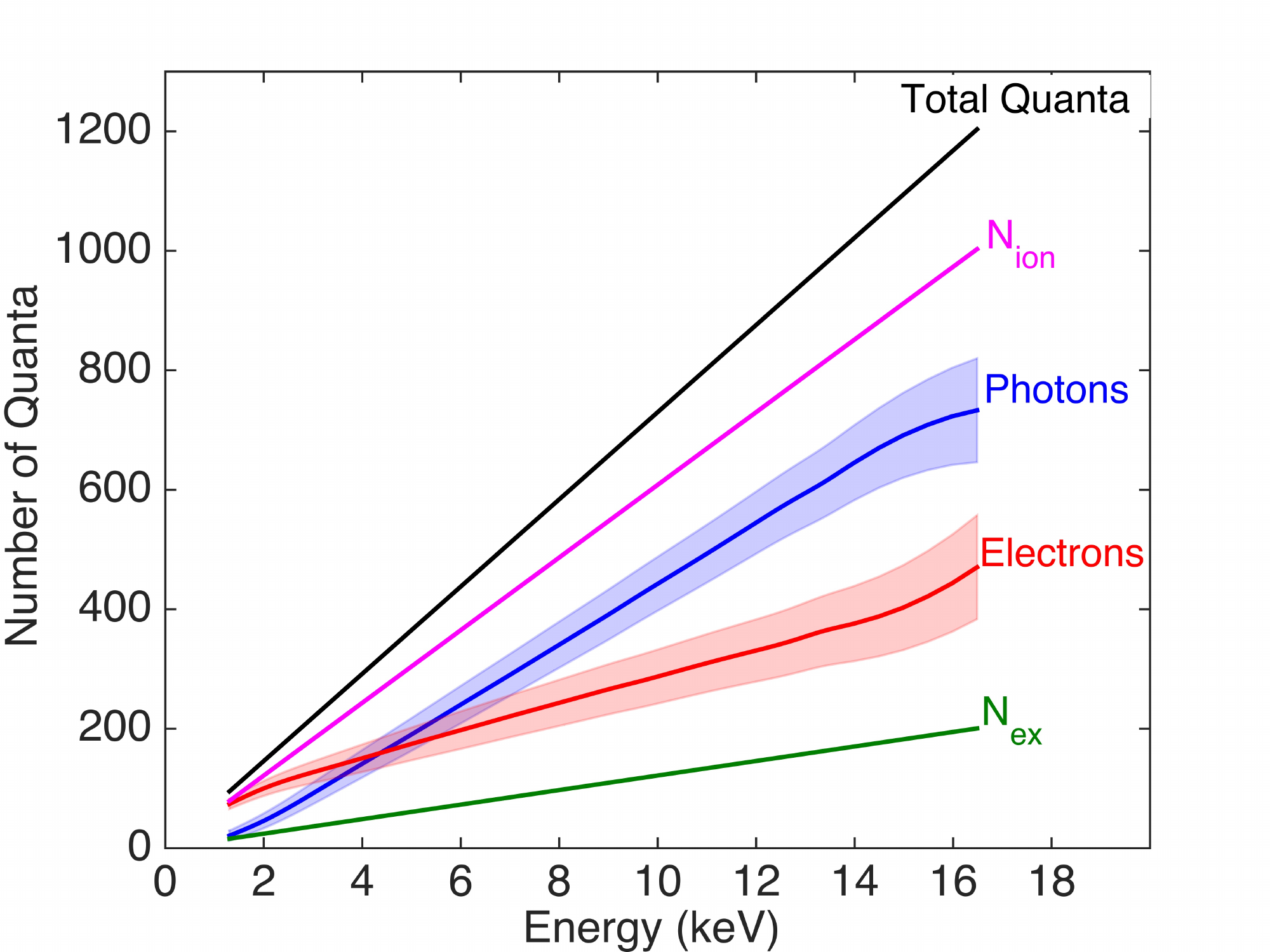}
\caption{Top: The mean number of electrons (red) and scintillation photons (blue) produced in LUX at 180~V/cm as a function of energy. The bands indicate the correlated systematic errors on $g_1$ and $g_2$. Also shown are the total number of quanta, primary ions, and primary excitons, assuming an exciton-to-ion ratio of $\alpha$ = 0.2. }
\label{fig:quanta-vs-energy}
\end{figure}

\subsection{Recombination at 180~V/cm and 105~V/cm}

As shown in Fig.~\ref{fig:ER-LY-QY}, we find that the light yield increases rapidly between 1~keV and 6~keV, and then becomes less energy-dependent over the remainder of the tritium spectrum, while the charge yield exhibits the complementary behavior. We understand these variations as being due to recombination, the process by which newly liberated ionization electrons are captured by Xe$^+$ ions, creating additional Xe$^*$ excitons, and ultimately scintillation photons~\cite{conti}. 

We model recombination as follows~\cite{Dahl_Thesis,Sorensen_Dahl,nest_2011}. Starting with a $W$-value of 13.7~eV, we assume that $\alpha$, the initial ratio of excitons-to-ions prior to recombination, is 0.2 independent of energy and electric field~\cite{Doke_alpha, Aprile_alpha}. Then the initial number of ions prior to recombination ($N_{ion}$, equivalent to the initial number of electrons), and the initial number of excitons prior to recombination ($N_{ex}$), and their sum (the total number of quanta), all increase linearly with energy as shown by the solid lines in Fig.~\ref{fig:quanta-vs-energy}. Also shown in Fig.~\ref{fig:quanta-vs-energy} are the total observed number of electrons and scintillation photons after recombination measured with the LUX tritium data at 180~V/cm as a function of energy. The sum of the observed electrons and photons should also increase linearly with energy, a hypothesis which is tested and confirmed by the tritium spectrum comparison of Fig.~\ref{fig:tritium-spectrum}.

\begin{figure}[t!]
\includegraphics[width=90mm]{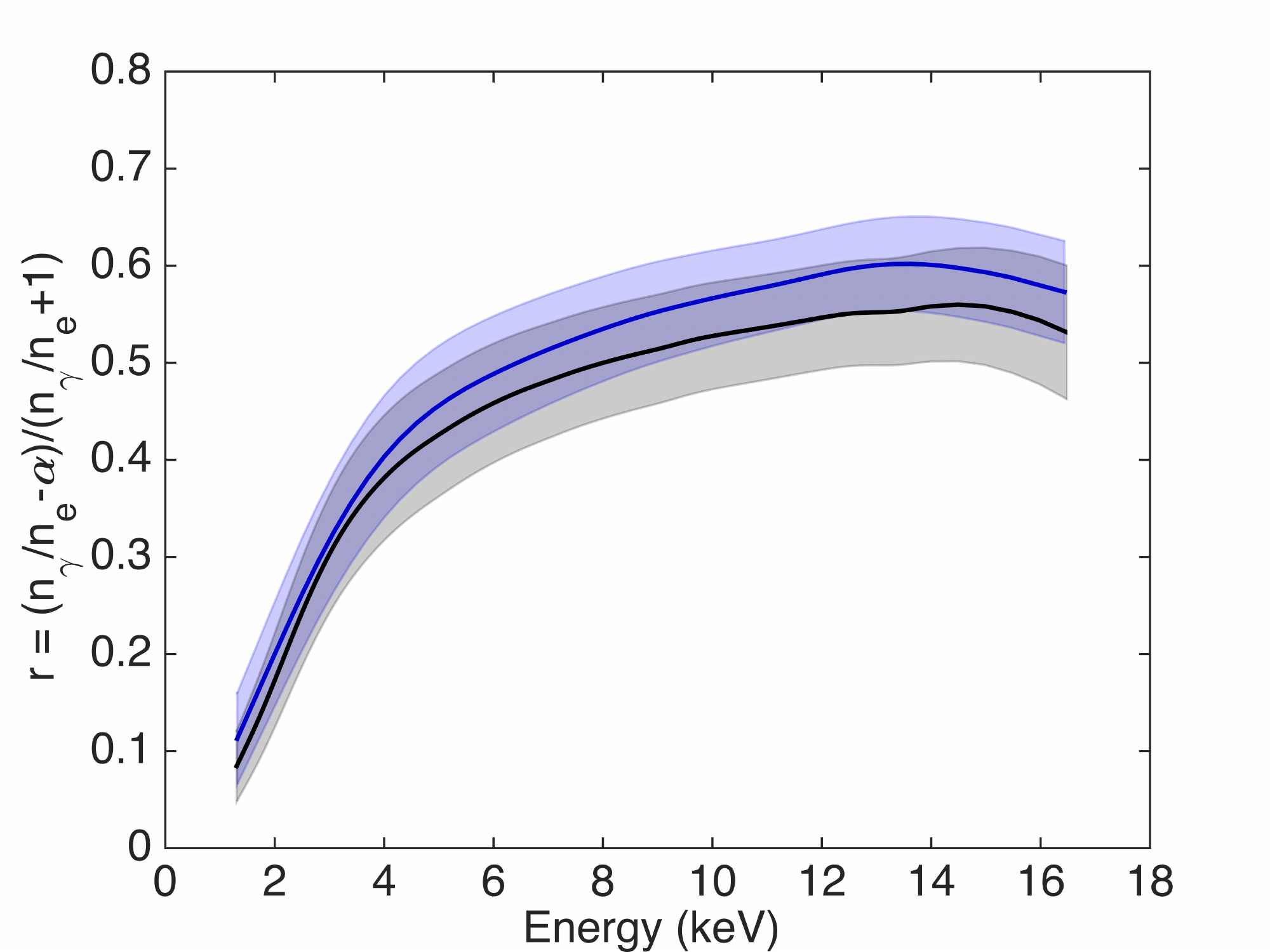}
\caption{Recombination fraction of ER events in LXe at 180~V/cm (black) and 105~V/cm (blue), assuming an exciton-to-ion ratio of 0.2.}
\label{fig:recombination}
\end{figure}

As shown in Fig.~\ref{fig:quanta-vs-energy}, we find that at very low energy, below 3~keV, the number of electrons and photons is similar to $N_{ion}$ and $N_{ex}$, respectively, while above 4~keV the number of electrons drops below the number of photons, consistent with a large recombination effect at these energies and this electric field. The recombination fraction, calculated according to

\begin{equation}
r = \frac{(n_{\gamma}/n_e) - \alpha}{(n_{\gamma}/n_e) + 1},
\end{equation}

\noindent
is shown explicitly in Fig.~\ref{fig:recombination}, measured with both the 180~V/cm and 105~V/cm tritium data. We find only a small difference in the recombination between these two field values in this energy range. It is worth noting that recombination is small at the very lowest energies where the dark matter search is performed, rapidly approaching zero as the energy drops below 4~keV. As noted before, this behavior is of considerable importance for the efficiency of recoil discrimination in LXe~\cite{xed-discrimination}. Other authors have used $\alpha$ values between 0.06 and 0.2 (see Ref.~\cite{kaixuan} and references therein). Changing the value of $\alpha$ modestly affects the absolute magnitude of the resulting recombination fraction but has only a small effect on the shape as a function of energy. 

\subsection{LUX electron recoil band}

\begin{figure}[t!]
\includegraphics[width=80mm]{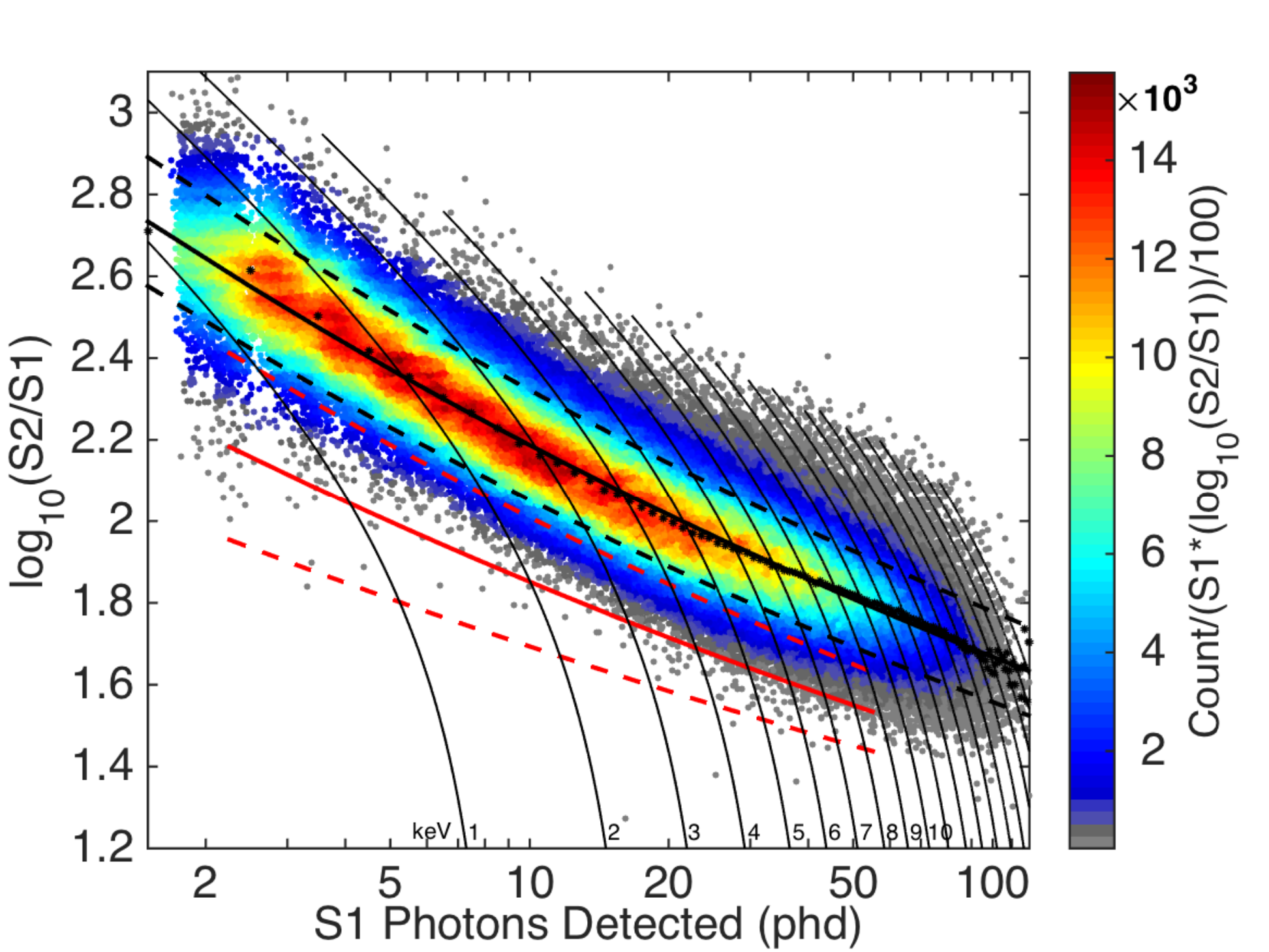} 
\includegraphics[width=75mm]{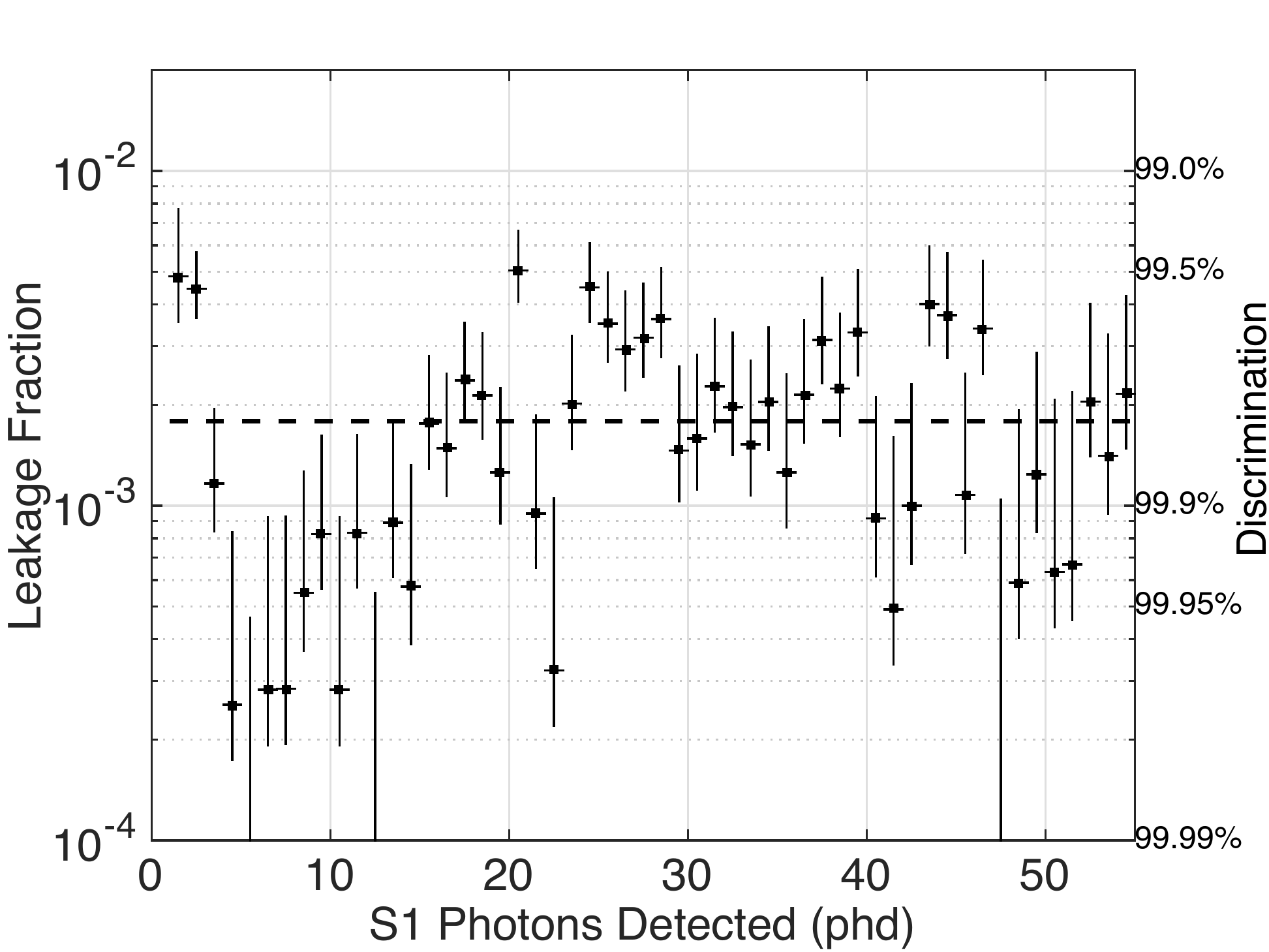}
\caption{Top: The electron recoil band of LUX illuminated by 170,000 tritium events at the nominal LUX electric field of 180~V/cm.  The recoil discriminant variable, $\rm log_{10}(S2/S1)$, is shown vs. $S1$ between 1 and 120 phd in S1 (with contours of constant ER energy from 1 to 20~keV). Also indicated in black are the Gaussian means in bins of S1 (filled dots), an empirical power law description of those means (solid black line), and the 10\% and 90\% contours of the ER population (dashed black lines). The solid red line represents the mean NR band determined with DD neutron generator data. The dashed red indicates the 10\% and 90\% contours of the NR band. Bottom: Observed leakage fraction vs. S1 between 1 and 50 phd. Y-axis labels: left: leakage fraction ($f$); right: discrimination ($1-f$).}
\label{fig:ER_band}
\end{figure}

The LUX ER band is shown as log$_{10}$(S2/S1) vs S1 in Fig.~\ref{fig:ER_band}(top).  It has a characteristic rise at decreasing values of S1 which reflects the rapidly changing charge and light yields below $\sim$6~keV.  Also shown in Fig.~\ref{fig:ER_band}(top) is the NR band measured with neutron generator data\cite{lux-reanalysis}. The width of the ER band is of considerable interest because it determines the recoil discrimination of the detector. The leakage fraction ($f$), defined as the fraction of ER events observed below the Gaussian mean of the NR band, is shown in Fig.~\ref{fig:ER_band}(bottom) as a function of S1. The recoil discrimination efficiency ($1-f$) has an average value of 99.81 \% $\pm$ 0.02\%$\rm (stat) \pm 0.1\%$(sys) for events with S1 between 1 and 50 phd, where the systematic error accounts for the uncertainty in the NR band mean and effects due to field non-uniformity.

\begin{figure}[t!]
\includegraphics[width=90mm]{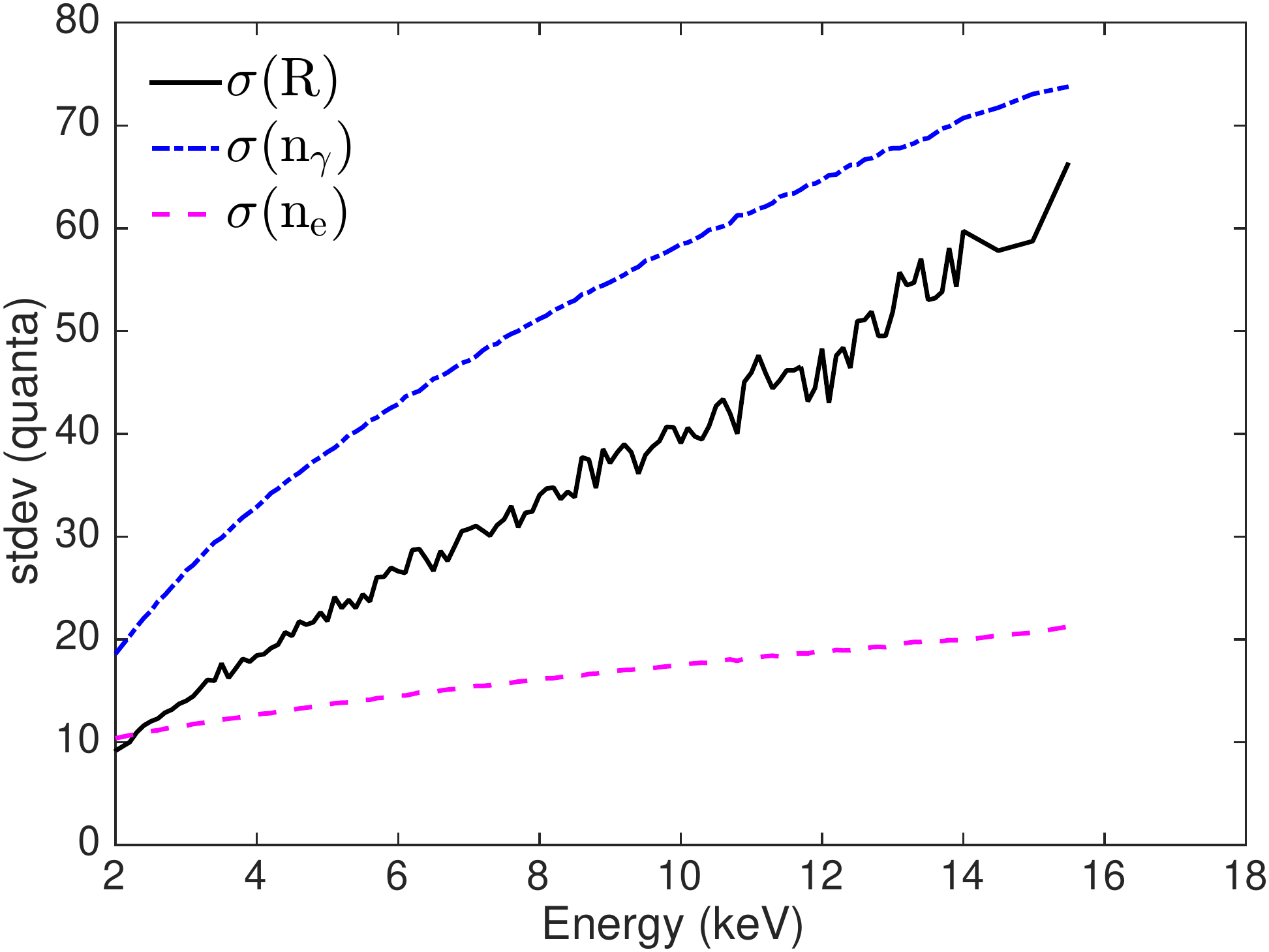}
\caption{Black: recombination fluctuations in LXe measured with LUX tritium data at 180~V/cm. Dot-dash blue: Detector resolution for counting photons. Dashed magenta: Detector resolution for counting electrons. }
\label{fig:recomb-flucs}
\end{figure}

In general the ER band width of an ideal detector should be comprised of three components: the uncertainties on photon counting and electron counting due to binomial collection statistics ($ \sigma(n_{\gamma})$ and $ \sigma(n_e)$), and the true event-to-event variations in recombination ($ \sigma(R)$). The binomial fluctuations are described by

\begin{eqnarray}
\sigma(n_{\gamma}) \sim \sqrt{(1-g_1)/(g_1*n_\gamma)}, \\
\sigma(n_{e}) \sim \sqrt{(1-\epsilon)/(\epsilon*n_e)},
\end{eqnarray}

\noindent
where $\epsilon$ is the electron extraction efficiency at the liquid surface. $ \sigma(n_{\gamma})$ and $ \sigma(n_e)$ also suffer additional variance due to PMT resolution, which can be measured with single photoelectron data.  Subtracting these sources of variance allows the recombination variance $ \sigma(R)$ to be isolated~\cite{Dobi_Thesis}. The method is cross-checked and confirmed with a toy Monte Carlo simulation where $ \sigma(n_{\gamma})$, $ \sigma(n_e)$, $ \sigma(R)$, and the PMT resolution are all known. The result for the LUX tritium data is shown in Fig.~\ref{fig:recomb-flucs} as a function of energy at 180~V/cm. The  recombination fluctuations are observed to grow linearly as a function of number of ions available for recombination. For energies between 2 to 16~keV the size of recombination fluctuations can be described by $\rm \sigma(R)=(0.067\pm0.005) \times N_{ion}$.

We find that at 180~V/cm in LUX, $ \sigma(n_{\gamma})$ is the most important contributor to the ER band width over the entire tritium energy spectrum due to the relatively modest light collection ($g_1 = 0.115$). Between 2 and 6~keV, where the WIMP search is most sensitive, $ \sigma(n_e)$ and $ \sigma(R)$ are of comparable magnitude and secondary importance. We note that an ideal detector, with perfect light and charge collection, would have an ER band width determined solely by $ \sigma(R)$.

\onecolumngrid
\break
\begin{figure}[t!]
\includegraphics[width=140mm]{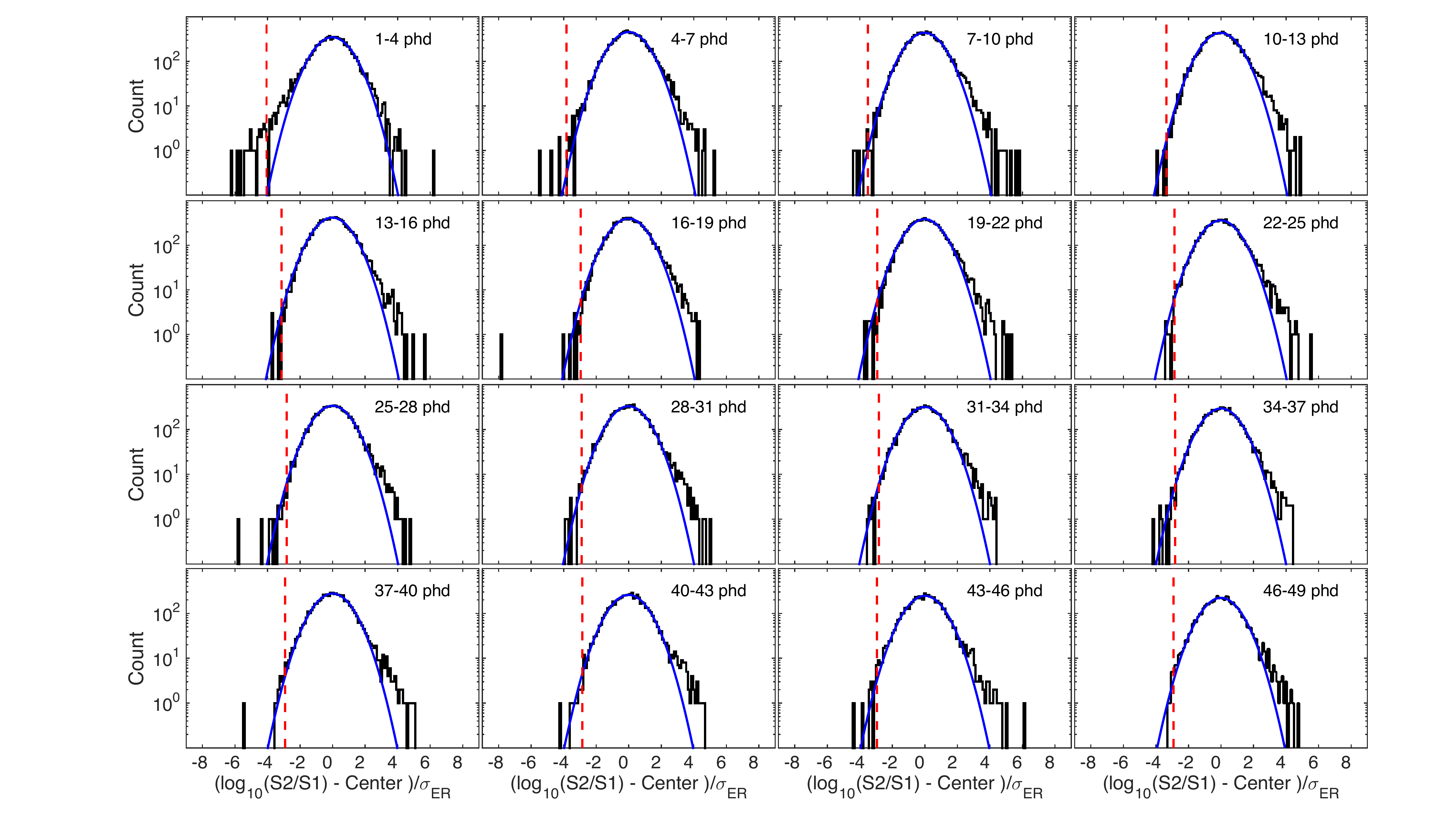}
\caption{Electron recoil population from tritium events in 3 phd bins over over the WIMP region of interest (1 -- 49 phd). We fit each bin to a Gaussian, and subtract the centroid of the Gaussian. The x-axis is measured in units of the fitted Gaussian width. The red dashed line represents the mean of the NR band in each bin. }
\label{fig:ER-Gauss}
\end{figure}
\twocolumngrid

The statistical description of the width of the LUX ER band is relevant to the WIMP-search profile likelihood fit. To study the band width in more detail, in Fig.~\ref{fig:ER-Gauss} we histogram log$_{10}(S2/S1)$ in 16 bins of S1 from 1 to 49 phd with a bin-width of 3 phd. In each bin, we show a Gaussian fit to the data after subtracting the centroid and dividing by the Gaussian width. We find that the Gaussian fits describe the data well  in most S1 bins out to $2\sigma$ on the upper side and $3\sigma$ on the lower side, beyond which non-Gaussian tails are visible.  We have investigated the origin of these tails. On the lower side, which is most directly relevant to the WIMP search, the largest non-Gaussian tail is found in the lowest S1 bin (1 -- 3 phd). This tail is reproduced in simulation and originates from Poissionian fluctuations in the photon counting statistics. The origin of the non-Gaussian tails on the upper side is less clear. It is worth noting that a similar effect has been seen in a previous experiment~\cite{zep3}.

Several outlier events are also evident in Fig.~\ref{fig:ER-Gauss}, particularly at low values of log$_{10}(S2/S1)$. Although these events are rare in this dataset, their origin is of considerable interest for understanding the WIMP sensitivity of future LXe experiments.  Therefore, we have investigated whether these events are attributable to detector pathologies, to backgrounds, or to the fundamental recombination physics of the LXe . In this dataset we expect to find about 0.5 low (S2/S1) events due to background ion recoil from $^{210}$Pb decay on the interior TPC walls. These events can have an improperly reconstructed radial position that allows them to pass our fiducial cuts.  
The $^{210}$Pb model is based upon a study of the WIMP search data and is described in Refs.~\cite{lux-reanalysis,Chang_Thesis}. Another possible background is from accidental coincidences between two distinct tritium events. In this scenario, an S1 from a tritium event below the cathode, and thus not having an S2, is improperly paired with a low energy tritium S2 in the fiducial volume for which the S1 signal fell below threshold. The S1 only rate during the tritium calibration is found by multiplying the total rate in the fiducial volume with the ratio of volume between the bottom PMT array and the cathode to the fiducial volume.  The S2 only rate is given by the total rate in the fiducial volume multiplied by the fraction of CH$_3$T events which fall below the S1 threshold of the detector.  An expectation of 2.5 accidental coincidence events in the tritium data is found by  multiplying the S1 only rate with the S2 only rate and integrating over the calibration live time and is found to be 2.5 events. The tritium dataset used here contain 27.5 live hours of data, during which time we expect to have 15 non-tritium events from the LUX ER background rate between 1 and 18 keV. These events should occur near the mean of the tritium ER band and should not be observable in this dataset. The total background expectation for low (S2/S1) events is therefore $\sim$3, and in Fig.~\ref{fig:ER-Gauss} we find three highly isolated low (S2/S1) events located in the 16-18, 25-28, and 37-40 phd bins. We conclude that the number of low (S2/S1) outlier events is consistent with the background expectation.

\section{Summary}

We have characterized the electron recoil response of the LUX dark matter experiment with a tritium calibration source. The large dataset, high event purity, and the single-site nature of the decay provide a powerful tool to study the detector and to investigate the fundamental properties of LXe as a particle detection medium for WIMP searches. 

We find strong evidence in support of the combined energy model for ER events in the WIMP energy range, and we report new measurements of the light and charge yields, the average recombination, and the fluctuations in the recombination as a function of energy. We have determined that the width of the ER band in LUX is driven by fluctuations in the number of detected S1 photons. We find a small number of outlier events far below the ER band centroid out of 170,000 fiducial tritium decays, consistent with background expectations in this dataset.

The results presented here are used in an improved analysis of the Run 3 WIMP search data to determine the location and width of the LUX ER band and to measure the fiducial volume~\cite{lux-reanalysis}. Additional tritium data has also been collected in support of the on-going LUX Run4 WIMP search and is presently under analysis. Furthermore, plans are being made to utilize a tritium source in the future LZ experiment~\cite{lz-cdr}, where external gamma sources such as $^{137}$Cs will produce a negligible rate of single scatter events in the fiducial region.

\appendix

\section{Studies of the removal of CH$_3$T from LXe }
\label{sec:appendix1}

\newcommand*{\Scale}[2][4]{\scalebox{#1}{$#2$}}%

Prior to the first injection of CH$_3$T into LUX, we considered three risks that such a calibration may pose to the dark matter search: 1) that the xenon purification system may be ineffective for CH$_3$T removal; 2) that the interior surfaces of the stainless steel (SS)  gas handling system may become permanently contaminated with CH$_3$T; and 3) that the plastic detector components may outgas unacceptable quantities of CH$_3$T after initial exposure.

\begin{figure}[h!]
\includegraphics[width=80mm]{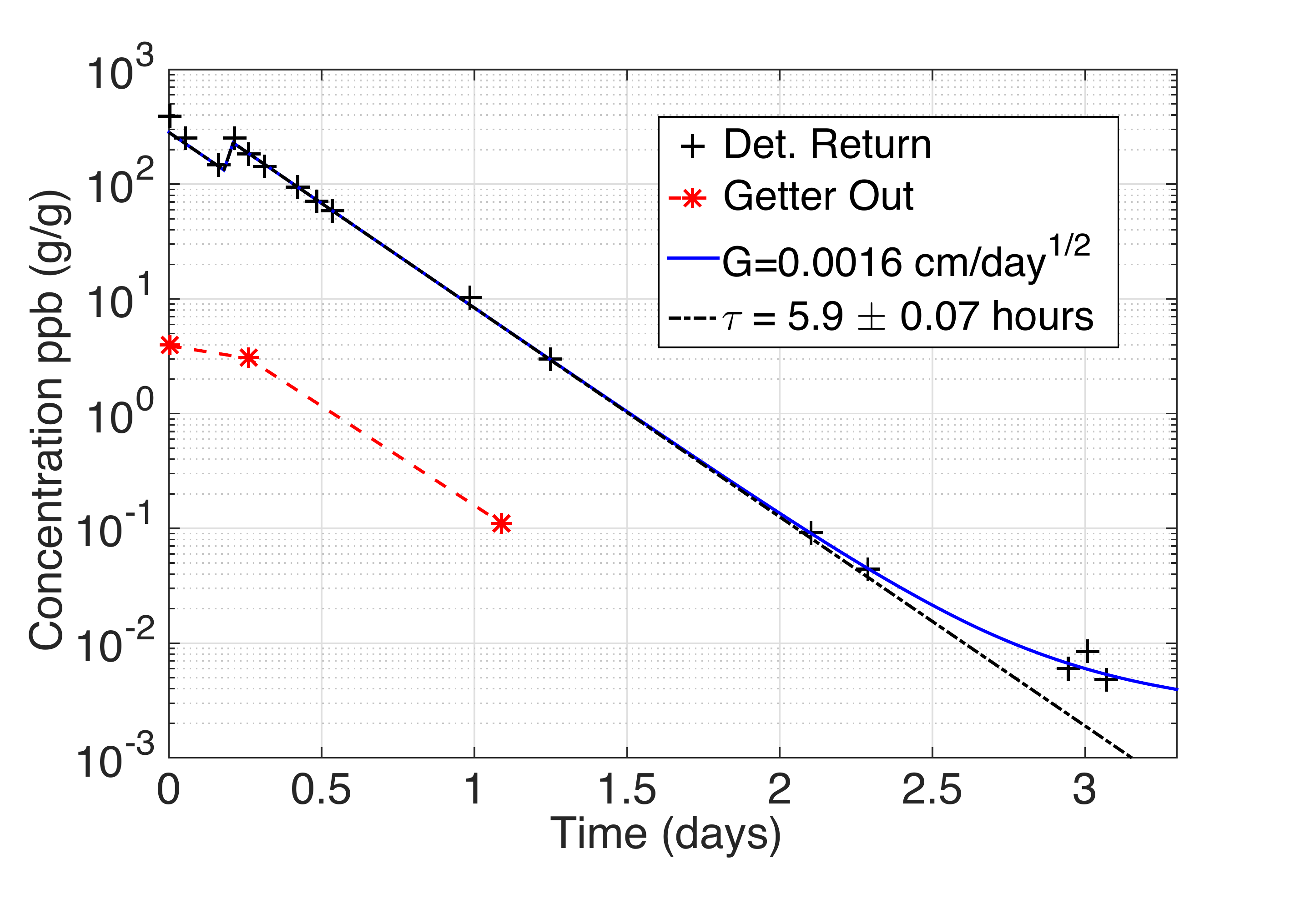}
\caption{Injection and removal of CH$_4$ in LUX prior to the first CH$_3$T injection. CH$_4$ is observed with a gas sampling mass spectrometry system. The black dashed lines shows an exponential fit to the CH$_4$ concentration at the detector return line with a time constant of 5.90 $\pm 0.07$ hours.  The red points indicate measurements at the getter outlet. We find a 97\% one-pass removal efficiency at a flow rate of 27 slpm. The blue curve shows the upper limit on the effect of outgassing from the plastics. The three data points near t = 3 days are consistent with the limit of detection for methane ($\sim \rm 5\times10^{-3}$ ppb (g/g)) .}
\label{fig:ch4_removal}
\end{figure}

To address the first concern we studied the removal of natural methane (CH$_4$) from Xe gas with a heated Zr getter and a mass spectrometer. The purification efficiency was found to be satisfactory~\cite{Dobi_CH4}. Furthermore, a test of the completed LUX purification system, including the actual getter unit, was performed several weeks before the first CH$_3$T injection into LUX. In this test $\sim$0.1 grams of CH$_4$ was injected into LUX, and mass spectrometry measurements of the CH$_4$ concentration in the LUX Xe gas were performed over the next several days. The CH$_4$ concentration was observed to decrease exponentially with a time constant of 5.90 $\pm 0.07$ hours as shown in Fig.~\ref{fig:ch4_removal}, confirming the effectiveness of the purification system for methane removal.

The behavior of CH$_3$T in SS plumbing was studied in a bench-test with a custom-built Xe gas proportional tube operated at room temperature. Substantial quantities of CH$_3$T activity were injected, counted, and removed from the proportional tube. Initial tests found a small amount of residual activity after purification, however this was resolved by passing the CH$_3$T through a methane purifier (SAES model MC1-905F). No subsequent contamination was observed.

\begin{figure}[h!]
\includegraphics[width=80mm]{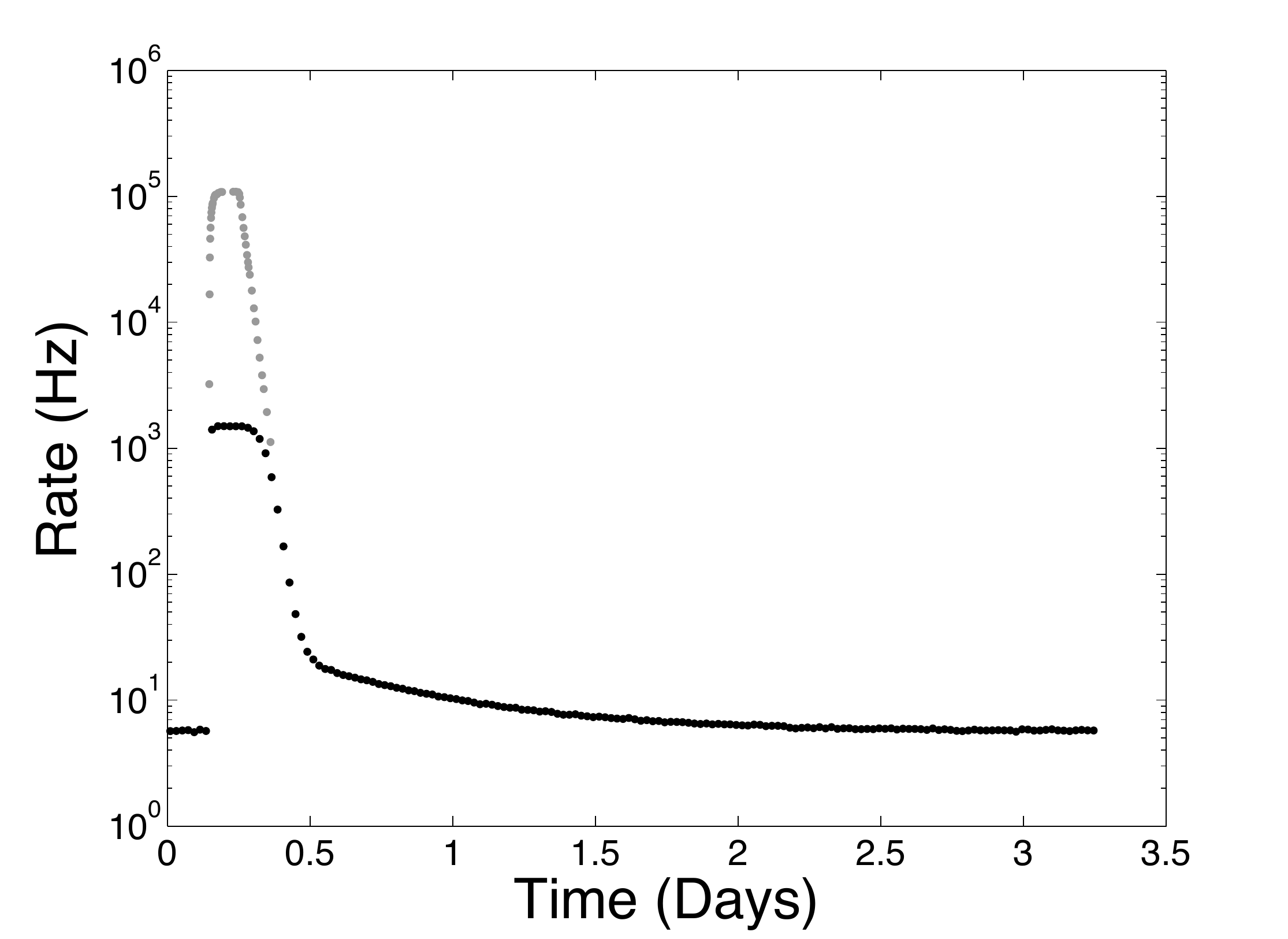}
\caption{The event rate versus time following a large CH$_3$T injection into a bench-top liquid xenon detector. Black points: the event rate measured with a dead-time limited digital DAQ system. Grey points: true event rate measured with a fast analog scalar. In this experiment a maximum activity of 68,000 Hz was detected immediately after the injection, compared to a background count rate of 5 Hz. Initially the purifier is not included in the recirculation circuit, leading to a constant count rate. The count rate falls rapidly when the purifier is activated. At 0.5 days an elbow in the count rate is observed, indicating that outgassing of CH$_3$T from the detector plastics has become a limiting factor in the purification rate. }
\label{fig:Density}
\end{figure}

We also performed tests of CH$_3$T injection and removal from LXe with a small detector. One such experiment is shown in Fig.~\ref{fig:Density}, where 68,000 Hz of CH$_3$T was injected, counted, and subsequently removed from LXe. Samples of LUX polyethylene and teflon were immersed in the LXe in this experiment, and their outgassing is evident in Fig.~\ref{fig:Density}. These data placed constraints on the risk of CH$_3$T outgassing in LUX. In total over one million Hz of CH$_3$T activity was injected and successfully removed in these experiments. 

As a final measure of risk mitigation, the CH$_3$T injection into LUX was performed at the end of Run 3 after the WIMP search data had been collected.

\section{Model of CH$_3$T removal}
\label{sec:appendix2}

We use a simple purification model to predict the CH$_3$T activity in LUX after an injection. The model is 

\begin{equation}
\frac{dC}{dt} = \frac{A}{V}J_{out} -\frac{C}{\tau},
\end{equation}

\noindent where  $C$ is the CH$_3$T concentration in the LXe,  $J_{out}$ is the flux of CH$_3$T out of the plastic components due to outgassing,  $A$ is the surface area of the plastic TPC cylinder, $V$ is the total volume of xenon in the active region, and $\tau$ is the characteristic removal time of CH$_3$T due to purification (5.9 hours). The model assumes perfect mixing of the fluid in the TPC, similar to what has been observed in LUX. The initial concentration is the injection activity divided by the volume of the active region. We solve the model numerically with the Euler method while simultaneously solving the diffusion equation to determine $J_{out}$. The results predict the number of calibration events that may be collected and provide an estimate of when the CH$_3$T  decay rate will be small enough to allow the WIMP search to resume.

We approximate the diffusion into and out of the plastics as one-dimensional, since most plastics in LUX can be approximated as a thin cylindrical shell with no dependence on the azimuthal or $z$ coordinates.  Fick's laws in one dimension are

\begin{align}
J=-D\frac{d \phi(r,t)}{d r}  \\
\frac{d \phi}{d t} = D \frac{d^2 \phi(r,t)}{d r^2} , 
\end{align}

\noindent 
where $J$ is the flux, $\phi(r,t)$ is the CH$_3$T  concentration in the plastic at depth $r$ and time $t$, and $D$ is the diffusion constant in the plastic. The concentration at the LXe-plastic boundary is fixed at $KC$, where K is the unitless solubility of CH$_3$T in the plastics. These equations are solved numerically and simultaneously with the purification model. 

$D$ and $K$ are not independently known for CH$_3$T in teflon or polyethylene at LXe temperature. However, only the combined quantity $G \equiv K \sqrt{ D/ \pi }$ is relevant as long as the diffusing substance does not reach the center of the plastic component (a good assumption for diffusion of CH$_3$T at LXe temperature). Under this condition, there exists an analytic solution to Fick's first law, which we evaluate at the LXe boundary:


\begin{equation}
J_{out}(t)= - G\left( \int \limits_0^t \frac{\frac{d}{dt'}C(t')}{\sqrt{t-t'}} dt' + \frac{C(0)}{\sqrt{t}}\right),
\end{equation}

\noindent
where the sign is reversed because the flux of material is outward. This result can be derived by applying Duhamel's principle along the infinite half-line, and it shows that the outgassing flux is linear in $G$. We set an upper limit of $G<0.0016 \frac{cm}{\sqrt{day}}$ for LUX based upon the data in Fig.~\ref{fig:ch4_removal}. In that data the effect of $G$ would appear as an elbow in the CH$_4$ concentration versus time, as indicated by the blue line. The three data points near t = 3 days constrain the maximum value of $G$. We interpret this result as an upper limit because those data points are consistent with CH$_4$ backgrounds in the mass spectrometry system.

\begin{figure}[h!]
\includegraphics[width=0.95\textwidth]{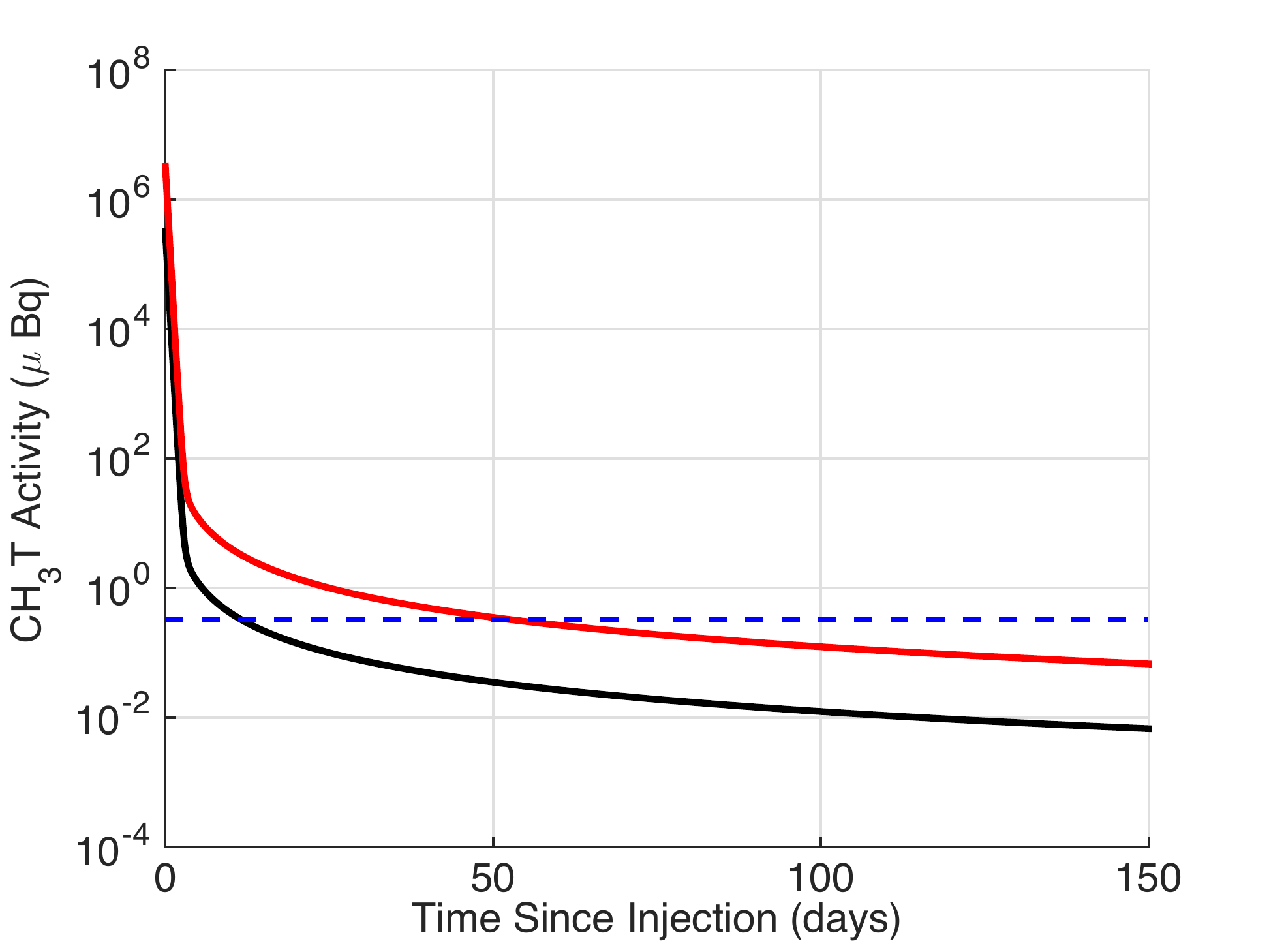}
\caption{Results of the purification model from 1 Bq (black curve) and 10 Bq (red curve) injections of CH$_3$T into LUX. The dashed blue line is the tritium activity goal of 0.33 $\mu$ Bq. The sharp initial fall is due to the 5.9 hour purification time constant of LUX, while the slow long-term removal is dominated by outgassing. The outgassing simulated here assumes  $G=0.0016$ cm/day$^{1/2}$).}
\label{fig:tau_var}
\end{figure}

Fig.~\ref{fig:tau_var} shows the results of the purification model for a 1 Bq and 10 Bq injection into LUX assuming $G$ = 0.0016 cm/day$^{1/2}$  . We take 0.33 $\mu$Bq of residual CH$_3$T activity as an approximate goal for resuming WIMP search running, and we find that for injections on the order of 1 Bq we reach 0.33 $\mu$Bq eight days later, while 10 Bq injections may take as long as 35 days.  Ultimately the final decision regarding low background data quality is made during the data analysis phase, with guidance provided by the purification model described here.

\section{Light and charge yields of electron recoils in LXe at 180~V/cm and 105~V/cm}
\label{sec:appendix3}

Tables~\ref{table:Yields} and \ref{table:Yields_100} list the light and charge yields of LXe for ER events between 1.3 and 17 keV and at fields of 180~V/cm and 105~V/cm, respectively. The uncertainties on the light and charge yields are highly anti-correlated in each energy bin due to the way in which the gain factors $g_1$ and $g_2$ are measured. The uncertainty listed includes both statistical and the dominant systematic uncertainty from the constraint on $g_1$ and $g_2$.

\begin{table}[h!]
\centering
\begin{tabular}{|c|c|c|c|} \hline
Energy 	& 		LY	& 	QY	& $\sigma$ \\ 
$\rm (keV_{ee}$) & ($\rm n_\gamma$/keV)   & ($\rm n_e$/keV) & ($\rm n$/keV) \\ \hline
1.3 	 & 14.6 	 & 58.4 	 & 2.2 	 \\ \hline 
1.5 	 & 17.3 	 & 55.7 	 & 1.9 	 \\ \hline 
2.0 	 & 22.3 	 & 50.7 	 & 2.4 	 \\ \hline 
2.5 	 & 27.4 	 & 45.6 	 & 2.5 	 \\ \hline 
3.0 	 & 31.5 	 & 41.4 	 & 2.3 	 \\ \hline 
3.5 	 & 33.8 	 & 39.2 	 & 2.0 	 \\ \hline 
4.0 	 & 35.8 	 & 37.2 	 & 2.2 	 \\ \hline 
4.5 	 & 37.5 	 & 35.5 	 & 2.0 	 \\ \hline 
5.0 	 & 38.4 	 & 34.6 	 & 1.9 	 \\ \hline 
5.2 	 & 38.9 	 & 34.1 	 & 2.0 	 \\ \hline 
5.5 	 & 39.5 	 & 33.5 	 & 2.1 	 \\ \hline 
6.0 	 & 40.4 	 & 32.6 	 & 2.0 	 \\ \hline 
6.5 	 & 41.7 	 & 31.3 	 & 2.0 	 \\ \hline 
7.0 	 & 41.7 	 & 31.3 	 & 1.7 	 \\ \hline 
7.5 	 & 42.7 	 & 30.3 	 & 2.0 	 \\ \hline 
8.0 	 & 42.9 	 & 30.1 	 & 1.9 	 \\ \hline 
9.0 	 & 43.8 	 & 29.1 	 & 1.7 	 \\ \hline 
10.0 	 & 44.7 	 & 28.3 	 & 2.0 	 \\ \hline 
11.0 	 & 45.4 	 & 27.6 	 & 1.7 	 \\ \hline 
12.0 	 & 46.0 	 & 27.0 	 & 1.7 	 \\ \hline 
13.0 	 & 46.5 	 & 26.5 	 & 1.5 	 \\ \hline 
14.0 	 & 47.1 	 & 25.9 	 & 1.6 	 \\ \hline 
16.0 	 & 46.4 	 & 26.6 	 & 2.5 	 \\ \hline 
17.0 	 & 44.9 	 & 28.1 	 & 2.5 	 \\ \hline 
\end{tabular}
\caption{Light and charge yield (photons/keV and electrons/keV) measured with tritium decay at 180~V/cm. The uncertainty includes both statistical and the dominant systematic uncertainty, common for both, from the constraint on $g_1$ and $g_2$.}
\label{table:Yields}
\end{table}

\begin{table}[h!]
\centering
\caption{Light and charge yield (photons/keV and electrons/keV) measured with tritium decay at 105~V/cm. The uncertainty includes both statistical and the dominant systematic uncertainty, common for both, from the constraint on $g_1$ and $g_2$. The available statistics in this data is smaller than that of the 180~V/cm data , resulting in fewer energy bins.}
\begin{tabular}{|c|c|c|c|} \hline
Energy 	& 		LY	& 	QY	& $\sigma$ \\ 
$\rm (keV_{ee}$) & ($\rm n_\gamma$/keV)   & ($\rm n_e$/keV) & ($\rm n$/keV) \\ \hline
1.3 	 & 18.4 	 & 54.6 	 & 1.7 	 \\ \hline 
2.2 	 & 25.1 	 & 47.8 	 & 1.9 	 \\ \hline 
3.1 	 & 33.4 	 & 39.6 	 & 2.2 	 \\ \hline 
4.0 	 & 37.6 	 & 35.4 	 & 2.3 	 \\ \hline 
4.9 	 & 39.9 	 & 33.1 	 & 1.7 	 \\ \hline 
5.8 	 & 41.3 	 & 31.6 	 & 2.2 	 \\ \hline 
6.7 	 & 43.0 	 & 29.9 	 & 2.0 	 \\ \hline 
7.6 	 & 44.1 	 & 28.9 	 & 1.6 	 \\ \hline 
8.5 	 & 46.2 	 & 26.8 	 & 2.0 	 \\ \hline 
9.4 	 & 46.2 	 & 26.8 	 & 2.0 	 \\ \hline 
10.3 	 & 47.7 	 & 25.3 	 & 1.5 	 \\ \hline 
11.2 	 & 46.8 	 & 26.2 	 & 1.5 	 \\ \hline 
12.1 	 & 49.1 	 & 23.9 	 & 1.9 	 \\ \hline 
13.0 	 & 49.6 	 & 23.4 	 & 1.5 	 \\ \hline 
13.9 	 & 50.8 	 & 22.2 	 & 3.2 	 \\ \hline 
15.0 	 & 49.2 	 & 23.7 	 & 1.4 	 \\ \hline 
16.4 	 & 46.3 	 & 26.7 	 & 2.2 	 \\ \hline 
\end{tabular}
\label{table:Yields_100}
\end{table}

\begin{acknowledgments}

This work was partially supported by the U.S. Department of Energy (DOE) under award numbers DE-FG02-08ER41549, DE-FG02-91ER40688, DE-FG02-95ER40917, DE-FG02-91ER40674, DE-NA0000979, DE-FG02-11ER41738, DE-SC0006605, DE-AC02-05CH11231, DE-AC52-07NA27344, and DE-FG01-91ER40618; the U.S. National Science Foundation under award numbers PHYS-0750671, PHY-0801536, PHY-1004661, PHY-1102470, PHY-1003660, PHY-1312561, PHY-1347449; the Research Corporation grant RA0350; the Center for Ultra-low Background Experiments in the Dakotas (CUBED); and the South Dakota School of Mines and Technology (SDSMT). LIP-Coimbra acknowledges funding from Funda\c{c}\~{a}o para a Ci\^{e}ncia e a Tecnologia (FCT)   through the project-grant PTDC/FIS-NUC/1525/2014. Imperial College and Brown University thank the UK Royal Society for travel funds under the International Exchange Scheme (IE120804). The UK groups acknowledge institutional support from Imperial College London, University College London and Edinburgh University, and from the Science \& Technology Facilities Council for PhD studentship ST/K502042/1 (AB). The University of Edinburgh is a charitable body, registered in Scotland, with registration number SC005336. This research was conducted using computational resources and services at the Center for Computation and Visualization, Brown University.

We gratefully acknowledge the logistical and technical support and the access to laboratory infrastructure provided to us by the Sanford Underground Research Facility (SURF) and its personnel at Lead, South Dakota. SURF was developed by the South Dakota Science and Technology Authority, with an important philanthropic donation from T. Denny Sanford, and is operated by Lawrence Berkeley National Laboratory for the Department of Energy, Office of High Energy Physics.

\end{acknowledgments}

\newpage
\thispagestyle{empty}
\mbox{}
\newpage
\thispagestyle{empty}
\mbox{}

\bibliography{Tritium}

\begin{thebibliography}{34}%
\makeatletter
\providecommand \@ifxundefined [1]{%
 \@ifx{#1\undefined}
}%
\providecommand \@ifnum [1]{%
 \ifnum #1\expandafter \@firstoftwo
 \else \expandafter \@secondoftwo
 \fi
}%
\providecommand \@ifx [1]{%
 \ifx #1\expandafter \@firstoftwo
 \else \expandafter \@secondoftwo
 \fi
}%
\providecommand \natexlab [1]{#1}%
\providecommand \enquote  [1]{``#1''}%
\providecommand \bibnamefont  [1]{#1}%
\providecommand \bibfnamefont [1]{#1}%
\providecommand \citenamefont [1]{#1}%
\providecommand \href@noop [0]{\@secondoftwo}%
\providecommand \href [0]{\begingroup \@sanitize@url \@href}%
\providecommand \@href[1]{\@@startlink{#1}\@@href}%
\providecommand \@@href[1]{\endgroup#1\@@endlink}%
\providecommand \@sanitize@url [0]{\catcode `\\12\catcode `\$12\catcode
  `\&12\catcode `\#12\catcode `\^12\catcode `\_12\catcode `\%12\relax}%
\providecommand \@@startlink[1]{}%
\providecommand \@@endlink[0]{}%
\providecommand \url  [0]{\begingroup\@sanitize@url \@url }%
\providecommand \@url [1]{\endgroup\@href {#1}{\urlprefix }}%
\providecommand \urlprefix  [0]{URL }%
\providecommand \Eprint [0]{\href }%
\providecommand \doibase [0]{http://dx.doi.org/}%
\providecommand \selectlanguage [0]{\@gobble}%
\providecommand \bibinfo  [0]{\@secondoftwo}%
\providecommand \bibfield  [0]{\@secondoftwo}%
\providecommand \translation [1]{[#1]}%
\providecommand \BibitemOpen [0]{}%
\providecommand \bibitemStop [0]{}%
\providecommand \bibitemNoStop [0]{.\EOS\space}%
\providecommand \EOS [0]{\spacefactor3000\relax}%
\providecommand \BibitemShut  [1]{\csname bibitem#1\endcsname}%
\let\auto@bib@innerbib\@empty
\bibitem [{\citenamefont {Akerib}\ \emph {et~al.}(2013)\citenamefont {Akerib}
  \emph {et~al.}}]{lux-nim}%
  \BibitemOpen
  \bibfield  {author} {\bibinfo {author} {\bibfnamefont {D.~S.}\ \bibnamefont
  {Akerib}} \emph {et~al.},\ }\href {\doibase 10.1016/j.nima.2012.11.135}
  {\bibfield  {journal} {\bibinfo  {journal} {Nucl.Instrum.Meth.}\ }\textbf
  {\bibinfo {volume} {A704}},\ \bibinfo {pages} {111} (\bibinfo {year}
  {2013})}\BibitemShut {NoStop}%
\bibitem [{\citenamefont {Akerib}\ \emph {et~al.}(2014)\citenamefont {Akerib}
  \emph {et~al.}}]{lux-prl}%
  \BibitemOpen
  \bibfield  {author} {\bibinfo {author} {\bibfnamefont {D.~S.}\ \bibnamefont
  {Akerib}} \emph {et~al.},\ }\href {\doibase 10.1103/PhysRevLett.112.091303}
  {\bibfield  {journal} {\bibinfo  {journal} {Phys.Rev.Lett.}\ }\textbf
  {\bibinfo {volume} {112}},\ \bibinfo {pages} {091303} (\bibinfo {year}
  {2014})}\BibitemShut {NoStop}%
\bibitem [{\citenamefont {Akerib}\ \emph
  {et~al.}(2015{\natexlab{a}})\citenamefont {Akerib} \emph
  {et~al.}}]{lux-reanalysis}%
  \BibitemOpen
  \bibfield  {author} {\bibinfo {author} {\bibfnamefont {D.~S.}\ \bibnamefont
  {Akerib}} \emph {et~al.} (\bibinfo {collaboration} {LUX}),\ }\href@noop {} {\
   (\bibinfo {year} {2015}{\natexlab{a}})},\ \bibinfo {note} {accepted for
  publication in Phys.Rev.Lett.},\ \Eprint {http://arxiv.org/abs/1512.03506}
  {arXiv:1512.03506 [astro-ph.CO]} \BibitemShut {NoStop}%
\bibitem [{\citenamefont {Kastens}\ \emph {et~al.}(2009)\citenamefont {Kastens}
  \emph {et~al.}}]{Kastens:2009pa}%
  \BibitemOpen
  \bibfield  {author} {\bibinfo {author} {\bibfnamefont {L.~W.}\ \bibnamefont
  {Kastens}} \emph {et~al.},\ }\href {\doibase 10.1103/PhysRevC.80.045809}
  {\bibfield  {journal} {\bibinfo  {journal} {Phys. Rev.}\ }\textbf {\bibinfo
  {volume} {C80}},\ \bibinfo {pages} {045809} (\bibinfo {year}
  {2009})}\BibitemShut {NoStop}%
\bibitem [{\citenamefont {Baudis}\ \emph {et~al.}(2013)\citenamefont {Baudis}
  \emph {et~al.}}]{Baudis}%
  \BibitemOpen
  \bibfield  {author} {\bibinfo {author} {\bibfnamefont {L.}~\bibnamefont
  {Baudis}} \emph {et~al.},\ }\href {\doibase 10.1103/PhysRevD.87.115015}
  {\bibfield  {journal} {\bibinfo  {journal} {Phys. Rev. D}\ }\textbf {\bibinfo
  {volume} {87}},\ \bibinfo {pages} {115015} (\bibinfo {year}
  {2013})}\BibitemShut {NoStop}%
\bibitem [{Note1()}]{Note1}%
  \BibitemOpen
  \bibinfo {note} {ER events and NR events generally have different energy
  scales in LXe. In this article we interpret all events using the ER energy
  scale.}\BibitemShut {Stop}%
\bibitem [{\citenamefont {Nagy}\ \emph {et~al.}(2006)\citenamefont {Nagy} \emph
  {et~al.}}]{Tritium_Q}%
  \BibitemOpen
  \bibfield  {author} {\bibinfo {author} {\bibfnamefont {S.}~\bibnamefont
  {Nagy}} \emph {et~al.},\ }\href@noop {} {\bibfield  {journal} {\bibinfo
  {journal} {Euro. Phys. Lett.}\ }\textbf {\bibinfo {volume} {74}},\ \bibinfo
  {pages} {404} (\bibinfo {year} {2006})}\BibitemShut {NoStop}%
\bibitem [{\citenamefont {Gregory}\ and\ \citenamefont
  {Landsman}(1958)}]{Tritium_Mean}%
  \BibitemOpen
  \bibfield  {author} {\bibinfo {author} {\bibfnamefont {D.~P.}\ \bibnamefont
  {Gregory}}\ and\ \bibinfo {author} {\bibfnamefont {D.~A.}\ \bibnamefont
  {Landsman}},\ }\href {\doibase 10.1103/PhysRev.109.2091} {\bibfield
  {journal} {\bibinfo  {journal} {Phys. Rev.}\ }\textbf {\bibinfo {volume}
  {109}},\ \bibinfo {pages} {2091} (\bibinfo {year} {1958})}\BibitemShut
  {NoStop}%
\bibitem [{\citenamefont {Venkataramaiah}\ \emph {et~al.}(1985)\citenamefont
  {Venkataramaiah} \emph {et~al.}}]{Tritium_Eq}%
  \BibitemOpen
  \bibfield  {author} {\bibinfo {author} {\bibfnamefont {P.}~\bibnamefont
  {Venkataramaiah}} \emph {et~al.},\ }\href
  {http://stacks.iop.org/0305-4616/11/i=3/a=014} {\bibfield  {journal}
  {\bibinfo  {journal} {Jour. Phys. G}\ }\textbf {\bibinfo {volume} {11}},\
  \bibinfo {pages} {359} (\bibinfo {year} {1985})}\BibitemShut {NoStop}%
\bibitem [{\citenamefont {Drexlin}\ \emph {et~al.}(2013)\citenamefont {Drexlin}
  \emph {et~al.}}]{Drexlin:2013lha}%
  \BibitemOpen
  \bibfield  {author} {\bibinfo {author} {\bibfnamefont {G.}~\bibnamefont
  {Drexlin}} \emph {et~al.},\ }\href {\doibase 10.1155/2013/293986} {\bibfield
  {journal} {\bibinfo  {journal} {Adv. High Energy Phys.}\ }\textbf {\bibinfo
  {volume} {2013}},\ \bibinfo {pages} {293986} (\bibinfo {year}
  {2013})}\BibitemShut {NoStop}%
\bibitem [{\citenamefont {Szydagis}\ \emph {et~al.}(2013)\citenamefont
  {Szydagis} \emph {et~al.}}]{NEST_2013}%
  \BibitemOpen
  \bibfield  {author} {\bibinfo {author} {\bibfnamefont {M.}~\bibnamefont
  {Szydagis}} \emph {et~al.},\ }\href {\doibase 10.1088/1748-0221/8/10/C10003}
  {\bibfield  {journal} {\bibinfo  {journal} {JINST}\ }\textbf {\bibinfo
  {volume} {8}},\ \bibinfo {pages} {C10003} (\bibinfo {year}
  {2013})}\BibitemShut {NoStop}%
\bibitem [{\citenamefont {Dobi}\ \emph {et~al.}(2010)\citenamefont {Dobi} \emph
  {et~al.}}]{Dobi_CH4}%
  \BibitemOpen
  \bibfield  {author} {\bibinfo {author} {\bibfnamefont {A.}~\bibnamefont
  {Dobi}} \emph {et~al.},\ }\href@noop {} {\bibfield  {journal} {\bibinfo
  {journal} {Nucl.Instrum.Meth.}\ }\textbf {\bibinfo {volume} {A620}},\
  \bibinfo {pages} {594 } (\bibinfo {year} {2010})}\BibitemShut {NoStop}%
\bibitem [{\citenamefont {Miyake}\ \emph {et~al.}(1983)\citenamefont {Miyake}
  \emph {et~al.}}]{miyake:1983}%
  \BibitemOpen
  \bibfield  {author} {\bibinfo {author} {\bibfnamefont {H.}~\bibnamefont
  {Miyake}} \emph {et~al.},\ }\href {\doibase 10.1116/1.572038} {\bibfield
  {journal} {\bibinfo  {journal} {Jour. Vac. Sci. Tech. A}\ }\textbf {\bibinfo
  {volume} {1}},\ \bibinfo {pages} {1447} (\bibinfo {year} {1983})}\BibitemShut
  {NoStop}%
\bibitem [{mor()}]{moravek}%
  \BibitemOpen
  \href@noop {} {}\bibinfo {note} {Moravek Biochemical Brea California 92821
  U.S.A}\BibitemShut {NoStop}%
\bibitem [{sea()}]{seas}%
  \BibitemOpen
  \href@noop {} {}\bibinfo {note} {SAES Pure Gas San Luis Obispo California
  93401 U.S.A.}\BibitemShut {Stop}%
\bibitem [{\citenamefont {Akerib}\ \emph
  {et~al.}(2015{\natexlab{b}})\citenamefont {Akerib} \emph {et~al.}}]{lux-prd}%
  \BibitemOpen
  \bibfield  {author} {\bibinfo {author} {\bibfnamefont {D.~S.}\ \bibnamefont
  {Akerib}} \emph {et~al.},\ }\href@noop {} {} (\bibinfo {year}
  {2015}{\natexlab{b}}),\ \bibinfo {note} {to be submitted to
  Phys.Rev.D}\BibitemShut {NoStop}%
\bibitem [{\citenamefont {Faham}\ \emph {et~al.}(2015)\citenamefont {Faham},
  \citenamefont {Gehman}, \citenamefont {Currie}, \citenamefont {Dobi},
  \citenamefont {Sorensen},\ and\ \citenamefont {Gaitskell}}]{doublepe}%
  \BibitemOpen
  \bibfield  {author} {\bibinfo {author} {\bibfnamefont {C.}~\bibnamefont
  {Faham}}, \bibinfo {author} {\bibfnamefont {V.}~\bibnamefont {Gehman}},
  \bibinfo {author} {\bibfnamefont {A.}~\bibnamefont {Currie}}, \bibinfo
  {author} {\bibfnamefont {A.}~\bibnamefont {Dobi}}, \bibinfo {author}
  {\bibfnamefont {P.}~\bibnamefont {Sorensen}}, \ and\ \bibinfo {author}
  {\bibfnamefont {R.}~\bibnamefont {Gaitskell}},\ }\href@noop {} {\bibfield
  {journal} {\bibinfo  {journal} {Journal of Instrumentation}\ }\textbf
  {\bibinfo {volume} {10}},\ \bibinfo {pages} {P09010} (\bibinfo {year}
  {2015})}\BibitemShut {NoStop}%
\bibitem [{\citenamefont {Platzman.}(1961)}]{Platzman}%
  \BibitemOpen
  \bibfield  {author} {\bibinfo {author} {\bibfnamefont {R.~L.}\ \bibnamefont
  {Platzman.}},\ }\href@noop {} {\bibfield  {journal} {\bibinfo  {journal}
  {Int.J.Appl.Radiat.Isotopes}\ }\textbf {\bibinfo {volume} {10}},\ \bibinfo
  {pages} {116} (\bibinfo {year} {1961})}\BibitemShut {NoStop}%
\bibitem [{\citenamefont {Dahl}(2009)}]{Dahl_Thesis}%
  \BibitemOpen
  \bibfield  {author} {\bibinfo {author} {\bibfnamefont {C.~E.}\ \bibnamefont
  {Dahl}},\ }\href@noop {} {Ph.D. thesis},\ \bibinfo  {school} {Princeton
  University} (\bibinfo {year} {2009})\BibitemShut {NoStop}%
\bibitem [{\citenamefont {Gushchin}\ \emph {et~al.}(1979)\citenamefont
  {Gushchin}, \citenamefont {Kruqlov}, \citenamefont {Litskevich},
  \citenamefont {Lebedev}, \citenamefont {Obodovskii},\ and\ \citenamefont
  {Somov}}]{gushchin:1979}%
  \BibitemOpen
  \bibfield  {author} {\bibinfo {author} {\bibfnamefont {E.}~\bibnamefont
  {Gushchin}}, \bibinfo {author} {\bibfnamefont {A.}~\bibnamefont {Kruqlov}},
  \bibinfo {author} {\bibfnamefont {V.}~\bibnamefont {Litskevich}}, \bibinfo
  {author} {\bibfnamefont {A.}~\bibnamefont {Lebedev}}, \bibinfo {author}
  {\bibfnamefont {I.}~\bibnamefont {Obodovskii}}, \ and\ \bibinfo {author}
  {\bibfnamefont {S.}~\bibnamefont {Somov}},\ }\href@noop {} {\bibfield
  {journal} {\bibinfo  {journal} {Sov. Phys. JETP}\ }\textbf {\bibinfo {volume}
  {49}},\ \bibinfo {pages} {856} (\bibinfo {year} {1979})}\BibitemShut
  {NoStop}%
\bibitem [{\citenamefont {Gushchin}\ \emph {et~al.}(1982)\citenamefont
  {Gushchin}, \citenamefont {Kruqlov},\ and\ \citenamefont
  {Obodovskii}}]{gushchin:1982}%
  \BibitemOpen
  \bibfield  {author} {\bibinfo {author} {\bibfnamefont {E.}~\bibnamefont
  {Gushchin}}, \bibinfo {author} {\bibfnamefont {A.}~\bibnamefont {Kruqlov}}, \
  and\ \bibinfo {author} {\bibfnamefont {I.}~\bibnamefont {Obodovskii}},\
  }\href@noop {} {\bibfield  {journal} {\bibinfo  {journal} {Sov. Phys. JETP}\
  }\textbf {\bibinfo {volume} {55}},\ \bibinfo {pages} {860} (\bibinfo {year}
  {1982})}\BibitemShut {NoStop}%
\bibitem [{\citenamefont {Dobi}(2014)}]{Dobi_Thesis}%
  \BibitemOpen
  \bibfield  {author} {\bibinfo {author} {\bibfnamefont {A.}~\bibnamefont
  {Dobi}},\ }\href@noop {} {Ph.D. thesis},\ \bibinfo  {school} {University of
  Maryland} (\bibinfo {year} {2014})\BibitemShut {NoStop}%
\bibitem [{Note2()}]{Note2}%
  \BibitemOpen
  \bibinfo {note} {We have verified with an internal $\protect \rm ^{83m}Kr$
  calibration source that the light yield of LXe is unaffected by the presence
  of CH$_4$ at concentrations up to $\sim $1 part per million. For the CH$_3$T
  measurements reported here the concentration was ($<$ $10\times 10^{-12}$
  g/g).}\BibitemShut {Stop}%
\bibitem [{\citenamefont {Aprile}\ \emph {et~al.}(2012)\citenamefont {Aprile}
  \emph {et~al.}}]{Aprile_LY}%
  \BibitemOpen
  \bibfield  {author} {\bibinfo {author} {\bibfnamefont {E.}~\bibnamefont
  {Aprile}} \emph {et~al.},\ }\href {\doibase 10.1103/PhysRevD.86.112004}
  {\bibfield  {journal} {\bibinfo  {journal} {Phys. Rev. D}\ }\textbf {\bibinfo
  {volume} {86}},\ \bibinfo {pages} {112004} (\bibinfo {year}
  {2012})}\BibitemShut {NoStop}%
\bibitem [{\citenamefont {Conti}\ \emph {et~al.}(2003)\citenamefont {Conti}
  \emph {et~al.}}]{conti}%
  \BibitemOpen
  \bibfield  {author} {\bibinfo {author} {\bibfnamefont {E.}~\bibnamefont
  {Conti}} \emph {et~al.},\ }\href {\doibase 10.1103/PhysRevB.68.054201}
  {\bibfield  {journal} {\bibinfo  {journal} {Phys. Rev. B}\ }\textbf {\bibinfo
  {volume} {68}},\ \bibinfo {pages} {054201} (\bibinfo {year}
  {2003})}\BibitemShut {NoStop}%
\bibitem [{\citenamefont {Sorensen}\ and\ \citenamefont
  {Dahl}(2011)}]{Sorensen_Dahl}%
  \BibitemOpen
  \bibfield  {author} {\bibinfo {author} {\bibfnamefont {P.}~\bibnamefont
  {Sorensen}}\ and\ \bibinfo {author} {\bibfnamefont {C.~E.}\ \bibnamefont
  {Dahl}},\ }\href {\doibase 10.1103/PhysRevD.83.063501} {\bibfield  {journal}
  {\bibinfo  {journal} {Phys. Rev. D}\ }\textbf {\bibinfo {volume} {83}},\
  \bibinfo {pages} {063501} (\bibinfo {year} {2011})}\BibitemShut {NoStop}%
\bibitem [{\citenamefont {Szydagis}\ \emph {et~al.}(2011)\citenamefont
  {Szydagis} \emph {et~al.}}]{nest_2011}%
  \BibitemOpen
  \bibfield  {author} {\bibinfo {author} {\bibfnamefont {M.}~\bibnamefont
  {Szydagis}} \emph {et~al.},\ }\href {\doibase 10.1088/1748-0221/6/10/P10002}
  {\bibfield  {journal} {\bibinfo  {journal} {JINST}\ }\textbf {\bibinfo
  {volume} {6}},\ \bibinfo {pages} {P10002} (\bibinfo {year}
  {2011})}\BibitemShut {NoStop}%
\bibitem [{\citenamefont {Doke}\ \emph {et~al.}(2002)\citenamefont {Doke} \emph
  {et~al.}}]{Doke_alpha}%
  \BibitemOpen
  \bibfield  {author} {\bibinfo {author} {\bibfnamefont {T.}~\bibnamefont
  {Doke}} \emph {et~al.},\ }\href
  {http://stacks.iop.org/1347-4065/41/i=3R/a=1538} {\bibfield  {journal}
  {\bibinfo  {journal} {Japan. Jour. Appl. Phys.}\ }\textbf {\bibinfo {volume}
  {41}},\ \bibinfo {pages} {1538} (\bibinfo {year} {2002})}\BibitemShut
  {NoStop}%
\bibitem [{\citenamefont {Aprile}\ \emph {et~al.}(2007)\citenamefont {Aprile}
  \emph {et~al.}}]{Aprile_alpha}%
  \BibitemOpen
  \bibfield  {author} {\bibinfo {author} {\bibfnamefont {E.}~\bibnamefont
  {Aprile}} \emph {et~al.},\ }\href {\doibase 10.1103/PhysRevB.76.014115}
  {\bibfield  {journal} {\bibinfo  {journal} {Phys. Rev. B}\ }\textbf {\bibinfo
  {volume} {76}},\ \bibinfo {pages} {014115} (\bibinfo {year}
  {2007})}\BibitemShut {NoStop}%
\bibitem [{\citenamefont {Shutt}\ \emph {et~al.}(2007)\citenamefont {Shutt}
  \emph {et~al.}}]{xed-discrimination}%
  \BibitemOpen
  \bibfield  {author} {\bibinfo {author} {\bibfnamefont {T.}~\bibnamefont
  {Shutt}} \emph {et~al.},\ }\href {\doibase 10.1016/j.nima.2007.04.104}
  {\bibfield  {journal} {\bibinfo  {journal} {Nucl.Instrum.Meth.}\ }\textbf
  {\bibinfo {volume} {A579}},\ \bibinfo {pages} {451} (\bibinfo {year}
  {2007})}\BibitemShut {NoStop}%
\bibitem [{\citenamefont {Lin}\ \emph {et~al.}(2015)\citenamefont {Lin} \emph
  {et~al.}}]{kaixuan}%
  \BibitemOpen
  \bibfield  {author} {\bibinfo {author} {\bibfnamefont {Q.}~\bibnamefont
  {Lin}} \emph {et~al.},\ }\href {\doibase 10.1103/PhysRevD.92.032005}
  {\bibfield  {journal} {\bibinfo  {journal} {Phys. Rev.}\ }\textbf {\bibinfo
  {volume} {D92}},\ \bibinfo {pages} {032005} (\bibinfo {year}
  {2015})}\BibitemShut {NoStop}%
\bibitem [{\citenamefont {Lebedenko}\ \emph {et~al.}(2009)\citenamefont
  {Lebedenko} \emph {et~al.}}]{zep3}%
  \BibitemOpen
  \bibfield  {author} {\bibinfo {author} {\bibfnamefont {V.~N.}\ \bibnamefont
  {Lebedenko}} \emph {et~al.},\ }\href {\doibase 10.1103/PhysRevD.80.052010}
  {\bibfield  {journal} {\bibinfo  {journal} {Phys. Rev.}\ }\textbf {\bibinfo
  {volume} {D80}},\ \bibinfo {pages} {052010} (\bibinfo {year}
  {2009})}\BibitemShut {NoStop}%
\bibitem [{\citenamefont {Lee}(2015)}]{Chang_Thesis}%
  \BibitemOpen
  \bibfield  {author} {\bibinfo {author} {\bibfnamefont {C.}~\bibnamefont
  {Lee}},\ }\href@noop {} {Ph.D. thesis},\ \bibinfo  {school} {Case Western
  Reserve University} (\bibinfo {year} {2015})\BibitemShut {NoStop}%
\bibitem [{\citenamefont {Akerib}\ \emph
  {et~al.}(2015{\natexlab{c}})\citenamefont {Akerib} \emph {et~al.}}]{lz-cdr}%
  \BibitemOpen
  \bibfield  {author} {\bibinfo {author} {\bibfnamefont {D.~S.}\ \bibnamefont
  {Akerib}} \emph {et~al.} (\bibinfo {collaboration} {LZ}),\ }\href@noop {} {\
  (\bibinfo {year} {2015}{\natexlab{c}})},\ \Eprint
  {http://arxiv.org/abs/1509.02910} {arXiv:1509.02910 [physics.ins-det]}
  \BibitemShut {NoStop}%
\end{thebibliography}%

\end{document}